\documentclass[a4paper, amsfonts, amssymb, amsmath, reprint, showkeys, footinbib, twoside,superscriptaddress,floatfix,longbibliography]{revtex4-1}

\usepackage{graphicx} 
\usepackage{amsmath}
\usepackage{comment}
\usepackage{booktabs}
\usepackage{hyperref}
\usepackage{url}
\usepackage{adjustbox}
\usepackage{array}
\usepackage[english]{babel}

\usepackage{multirow}
\usepackage{hhline}
\usepackage{makecell}
\usepackage[table]{xcolor}

\newcommand{\figLabelCapt}[1]{\textbf{\MakeLowercase{{#1}}}}
\newcommand{\refSub}[2]{\hyperref[#2]{\ref{#2}\figLabelCapt{#1}}}

\newcommand{\br}[1]{\mathbf{r}}
\newcommand{\bk}[1]{\mathbf{k}}

\begin{document}

\title{Long-range electrostatics for machine learning interatomic potentials \\is easier than we thought}

\author{Dongjin Kim}
\affiliation{Department of Chemistry, UC Berkeley, California 94720, United States}

\author{Bingqing Cheng}
\email{bingqingcheng@berkeley.edu}
\affiliation{Department of Chemistry, UC Berkeley, California 94720, United States}
\affiliation{Chemical Sciences Division, Lawrence Berkeley National Laboratory, Berkeley, California, 94720, United States}
\affiliation{Bakar Institute of Digital Materials for the Planet, UC Berkeley, California 94720, United States}
\affiliation{The Institute of Science and Technology Austria, Am Campus 1, 3400 Klosterneuburg, Austria}

\date{\today}

\begin{abstract}
The lack of long-range electrostatics is a key limitation of modern machine learning interatomic potentials (MLIPs), hindering reliable applications to interfaces, charge-transfer reactions, polar and ionic materials, and biomolecules. In this Perspective, we distill two design principles behind the Latent Ewald Summation (LES) framework, which can capture long-range interactions, charges, and electrical response just by learning from standard energy and force training data: (i) use a Coulomb functional form with environment-dependent charges to capture electrostatic interactions, and (ii) avoid explicit training on ambiguous density functional theory (DFT) partial charges. When both principles are satisfied, substantial flexibility remains: essentially any short-range MLIP can be augmented; charge equilibration schemes can be added when desired; dipoles and Born effective charges can be inferred or finetuned; and charge/spin-state embeddings or tensorial targets can be further incorporated. We also discuss current limitations and open challenges. Together, these minimal, physics-guided design rules suggest that incorporating long-range electrostatics into MLIPs is simpler and perhaps more broadly applicable than is commonly assumed.
\end{abstract}

\maketitle


There is no denying that accurate modeling of the $1/r$-decaying long-range electrostatics is crucial in atomistic simulations, particularly for interfaces, charge-transfer processes, polar and ionic materials, and biomolecular systems~\cite{french2010long, York1995Accurate, rodgers2008interplay}. 
The open question has always been how ``easy'' it is to incorporate such long-range interactions into the machine learning interatomic potential (MLIP) framework.

Most MLIPs are built on the locality assumption~\cite{prodan2005nearsightedness}: the energy and forces associated with a central atom are largely determined by its local neighbors~\cite{keith2021combining, unke2021machine}. 
The total short-range energy of a system is thus approximated as a sum of local atomic energy contributions $E_i$ from each atom $i$, $E^{\mathrm{sr}}=\sum_{i=1}^{N} E_i$. 
Early MLIPs~\cite{behler2007generalized, bartok2010gaussian} define the local environment using a fixed spatial cutoff radius, typically on the order of 6~\AA{}.
Later message-passing neural networks (MPNNs)~\cite{schutt2017schnet, haghighatlari2022NewtonNet, batzner20223, batatia2022mace, fu2025Learning, wood2025umafamilyuniversalmodels} propagate information across multiple interaction shells, effectively increasing the receptive field. 
However, this is no guarantee that long-range interactions are faithfully captured, as deep MPNNs can suffer from information degradation and over-smoothing~\cite{Li2018Deeper, Alon2021Bottleneck}.

\begin{figure*}[t]
\centering
\includegraphics[width=0.7\linewidth]{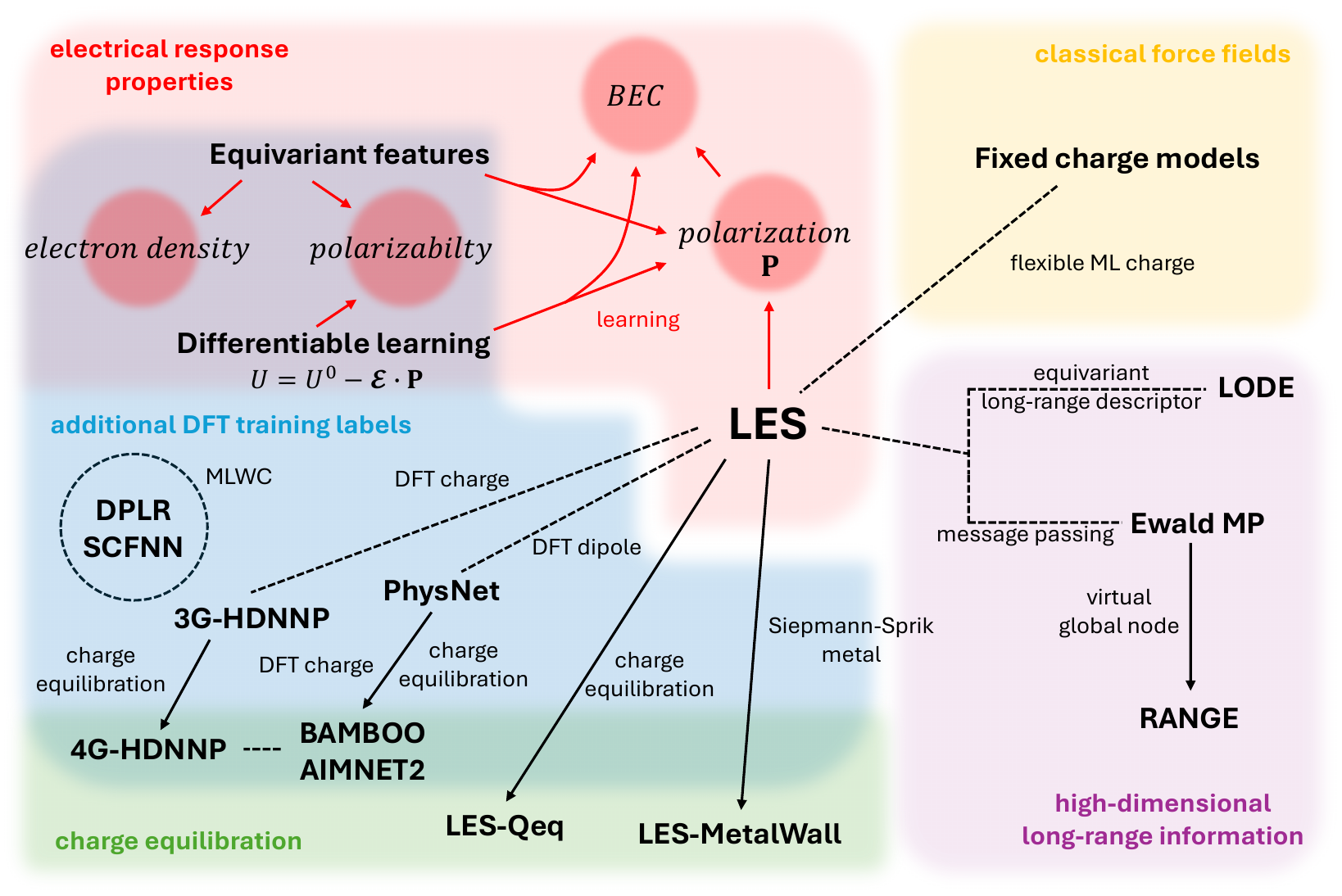}
     \caption{
     A map of representative strategies for learning long-range interactions.
     Black solid arrows indicate an extension or variant of a method. 
     Black dashed arrows show conceptual connections. 
     Red arrows indicate the learning of physical quantities.
     The blue shaded region indicates methods that use additional DFT training labels, the green denotes methods with charge equilibration, the purple marks the methods using high-dimensional long-range representations, the yellow represents the classical force fields with fixed charges, and the red shows the methods that can learn electrical response properties.
     Abbreviations: LES, latent Ewald summation~\cite{Cheng2025Latent}; HDNNP, high-dimensional neural network potential~\cite{Morawietz2012neural, ko2021fourth}; BAMBOO, ByteDance AI molecular simulation booster~\cite{gong_predictive_2025}; AIMNET, atoms-in-molecules neural network potential~\cite{Anstine2025}; Ewald MP, Ewald message passing; LODE, long-distance equivariant~\cite{grisafi2019incorporating}; RANGE, relaying attention nodes for global encoding~\cite{caruso2025Extending}; DPLR, deep potential long-range~\cite{zhang2022deep}; SCFNN, self-consistent field neural network~\cite{gao2022self}; Qeq, charge equilibration scheme proposed in Ref.~\cite{rappe1991charge, Ghasemi2015Interatomic}; BEC, Born effective charge tensors;
     Equivariant features, direct prediction of the tensorial quantities~\cite{grisafi2018symmetry, schienbein2023Spectroscopy, Grisafi2019Transferable}; 
     Differentiable learning, learn corresponding response properties by taking derivatives~\cite{falletta2025unified,Schmiedmayer2024}.
     $U$ is electric enthalpy, $U^0$ is the energy in the absence of the electric field, $P$ is polarization, and $\mathcal{E}$ is a uniform electric field~\cite{falletta2025unified}.
    }
    \label{fig:lr_map}
\end{figure*}

In recent years, various strategies~\cite{Ko2021GeneralPurpose, unke2021machine, Kocer2022Neural, Behler2021Four, Anstine2023Machinea, Cheng2025Latent, unke2019physnet, ko2021fourth, sifain2018discovering, gong_predictive_2025, shaidu2024incorporating, Song2024ChargeOptimizeda, Kabylda2025, Jacobson2022Transferablea, Thurlemann2022Learninga, Morawietz2012neural, Anstine2025, Rinaldi2025Chargeconstrained, unke2021spookynet, zhang2022deep, gao2022self, yu2022capturing, kosmala2023ewald, grisafi2019incorporating, huguenin2023physics, Faller2024Densitybased,monacelli2024electrostatic, Loche2025Fastb, Rumiantsev2025Learning, caruso2025Extending, weber2025efficient, Ghasemi2015Interatomic, fuchs2025Learninga, Maruf2025Learninga, Xu2025universala, ple2025Foundation, Gao2025foundation}
have been proposed to combine electrostatics with MLIPs.
In Fig.~\ref{fig:lr_map}, we illustrate the representative developments and the conceptual connections between them. This is by no means exhaustive and is skewed towards the earlier methods in each category, particularly those that have inspired our own work.

The methods in the purple shaded region of Fig.~\ref{fig:lr_map} capture long-range effects by embedding global geometrical information directly into the model representation~\cite{kosmala2023ewald, caruso2025Extending, grisafi2019incorporating, huguenin2023physics, Faller2024Densitybased,monacelli2024electrostatic, yu2022capturing, Rumiantsev2025Learning}. 
This includes approaches based on global or reciprocal-space message passing~\cite{kosmala2023ewald, Rumiantsev2025Learning},
the use of virtual global nodes for communicating global information~\cite{caruso2025Extending}, 
and those constructing explicit long-range descriptors~\cite{grisafi2019incorporating, Faller2024Densitybased, huguenin2023physics}. 
In these frameworks, long-range interactions are not introduced as a physical electrostatic energy term, but rather as part of the learned global feature space from which the neural network can infer nonlocal interactions.

The methods inside the blue shaded region of Fig.~\ref{fig:lr_map} encode physically-motivated quantities such as atomic charges~\cite{Morawietz2012neural, ko2021fourth, unke2019physnet, gong_predictive_2025, Anstine2025, Rinaldi2025Chargeconstrained, sifain2018discovering, Kabylda2025, Song2024ChargeOptimizeda, Jacobson2022Transferablea, unke2021spookynet,Maruf2025Learninga, fuchs2025Learninga}, typically trained on electronic information from density functional theory (DFT).
Notably, 3G-HDNNP~\cite{Morawietz2012neural,ko2021fourth} learns DFT partial charges~\cite{Morawietz2012neural}, and this was later refined in 4G-HDNNP~\cite{ko2021fourth} that further include charge equilibration (Qeq) scheme~\cite{rappe1991charge, Ghasemi2015Interatomic}.
PhysNet~\cite{unke2019physnet}, BAMBOO~\cite{gong_predictive_2025}, and AIMNET2~\cite{Anstine2025} include DFT dipoles in the loss function and predict atomic partial charges.
Deep potential long-range (DPLR) model~\cite{zhang2022deep} and the self-consistent field neural network (SCFNN)~\cite{gao2022self} learn from the maximally localized Wannier centers (MLWCs) for insulating systems.

In a separate vein, the methods highlighted in the red shaded region specifically target electrical response properties such as polarization ($P$)~\cite{falletta2025unified, Schmiedmayer2024, Stocco2025Electricfield, Zhang2023, Gastegger2021, King2025Machine, zhong2025machine, Kim2025Universalb}, polarizability~\cite{falletta2025unified, Zhang2023, Gastegger2021}, electron density~\cite{Grisafi2019Transferable, Lewis2021Learning, grisafi2023predicting}, and Born effective charge (BEC) tensors ($Z^*$)~\cite{zhong2025machine, King2025Machine, Kim2025Universalb, falletta2025unified, yu_prediction_2025, Schmiedmayer2024, Stocco2025Electricfield}. 
These quantities may be learned directly as tensorial outputs using equivariant features~\cite{grisafi2018symmetry, Schmiedmayer2024, schienbein2023Spectroscopy, Joll2024, yu_prediction_2025, Stocco2025Electricfield, Grisafi2019Transferable, Lewis2021Learning}, or through differentiable learning frameworks in which response properties are obtained as derivatives of learned quantities~\cite{falletta2025unified, zhong2025machine}, for examples, BECs as derivatives of polarization with respect to atomic displacements~\cite{falletta2025unified, Stocco2025Electricfield, zhong2025machine}, and polarizations $P$ as derivatives of energy with respect to the external electric field $\mathcal{E}$~\cite{falletta2025unified, Gastegger2021, Zhang2023}.
Related developments by Grisafi et al.~\cite{Grisafi2019Transferable, grisafi2023predicting} introduced long-range symmetry-adapted models for predicting electron densities and the electronic density response under applied fields.

Latent Ewald Summation (LES)~\cite{Cheng2025Latent, King2025Machine, zhong2025machine, Kim2025Universalb} is at the intersection of the three aforementioned approaches: 
LES explicitly employs atomic charges to construct a physical long-range Coulomb electrostatic energy term; however, these charges are not learned from DFT electronic labels. 
Instead, they are inferred as latent, environment-dependent features solely from training on standard energy and force data.
Meanwhile, electrical response properties such as  $P$ and $Z^*$ emerge naturally~\cite{King2025Machine, zhong2025machine, Kim2025Universalb}.

The LES algorithm is simple: 
the local invariant features on each atom $i$ (same as the ones for predicting local energies $E_i$) predict the latent atomic charge $q^\mathrm{les}$ via a neural network.
These $q^\mathrm{les}$ are then used to compute
long-range electrostatic energy $E^\mathrm{lr}$:
Pairwise summation is used for finite systems,
\begin{equation}
    E^{\mathrm{lr}}= \frac{1}{2}\frac{1}{4\pi \varepsilon_0} \sum_{i=1}^{N} \sum_{j=1}^N   [1 - \varphi(r_{ij})] \dfrac{q^\mathrm{les}_i q^\mathrm{les}_j}{r_{ij}},
    \label{eq:e_lr-finite}
\end{equation}
where $\varepsilon_0$ is the vacuum permittivity, $\varphi(r) = \mathrm{erfc}(\frac{r}{\sqrt{2}\sigma})$ is the complementary error function, $\sigma$ is a smearing factor typically $\approx1$~\AA.
Ewald summation is used for periodic systems, 
\begin{equation}
E^\mathrm{lr} = \dfrac{1}{2\varepsilon_0V} \sum_{0<k<k_c} \dfrac{1}{k^2} e^{-\sigma^2 k^2/2} |S(\mathbf{k})|^2,
\label{eq:e_lr}
\end{equation}
with
\begin{equation}
S(\mathbf{k}) = \sum_{i=1}^N q_i^\mathrm{les} e^{i\mathbf{k}\cdot\mathbf{r}_i},
\label{eq:sfactor}
\end{equation}
where $\mathbf{k}$ is the reciprocal wave vector, and 
$V$ is the cell volume.
Finally, the electrostatic energy is added to the short-range energy to form the total potential energy of the system, $E=E^\mathrm{sr}+E^\mathrm{lr}$.

For a finite system, the polarization is $P= \sum_{i=1}^N q_i \mathbf{r}_i$, and the BEC for an atom $i$ is
\begin{equation}
Z^*_{i \alpha\beta} 
= \dfrac{\partial P_{\alpha}}{\partial r_{i\beta}},
\label{eq:z-finite}
\end{equation}
with $\alpha$ and $\beta$ labeling Cartesian directions.
For a homogeneous periodic system,
\begin{equation}
Z^*_{i \alpha\beta} 
= \lim_{k\rightarrow0}\Re\left[\exp(-ik r_{i\alpha}) \dfrac{\partial P_{\alpha} (k)}
{\partial r_{i\beta}}\right],
\label{eq:z-pbc}
\end{equation}
where $P_\alpha(k) = \sum_{i=1}^{N} \sqrt{\varepsilon_\infty}\dfrac{q_i^\mathrm{les}}{ik} \exp(i k r_{i\alpha}) $,
and $\varepsilon_\infty$ is high-frequency (electronic) relative permittivity.

Besides the simple algorithm, the implementation of LES into any short-range MLIP is also easy.
Thus far, LES has been integrated with MACE~\cite{batatia2022mace}, NequIP~\cite{batzner20223}, Allegro~\cite{musaelian2023learning}, CACE~\cite{cheng2024cartesian}, CHGNet~\cite{deng2023chgnet}, and these implementations are discussed and benchmarked in Ref.~\cite{Kim2025Universalb}.
In addition, LES has been independently implemented in TACE~\cite{Xu2025universala}.
In Refs.~\cite{Cheng2025Latent, King2025Machine, zhong2025machine, Kim2025Universalb, Wang2025Ionmodulateda}, LES has been extensively benchmarked on diverse systems, including bulk liquids~\cite{Cheng2025Latent, King2025Machine, zhong2025machine, Kim2025Universalb}, molecular crystals~\cite{Kim2025Universalb}, polar molecules~\cite{Cheng2025Latent, King2025Machine, Kim2025Universalb}, ferroelectrics~\cite{zhong2025machine}, and electrochemical interfaces~\cite{King2025Machine, Wang2025Ionmodulateda}. 
Overall, the LES augmentation was shown to universally improve the accuracy and ensure the reliable extraction of electrical response properties, including BECs~\cite{King2025Machine, zhong2025machine, Kim2025Universalb}, dipoles for finite systems~\cite{King2025Machine, Kim2025Universalb}, infrared spectra~\cite{zhong2025machine, Kim2025Universalb}, and ionic electrical conductivities~\cite{zhong2025machine}.
Meanwhile, adding LES only introduces a marginal computational overhead~\cite{King2025Machine, Kim2025Universalb}, making it effectively a ``free lunch''.

\begin{figure*}
    \centering
    \includegraphics[width=0.9\linewidth]{fig2.pdf}
    \caption{Benchmark results of short-range and LES-augmented MLIPs for bulk RPBE-D3 water.
    \figLabelCapt{a}:
    The test force root mean square errors (RMSEs) for baseline MLIPs (hollow bars), and the corresponding LES-augmented models (solid bars) for different architectures (MACE~\cite{batatia2022mace}, NequIP~\cite{batzner20223}, CHGNet~\cite{deng2023chgnet}, CACE~\cite{cheng2024cartesian}, and Allegro~\cite{musaelian2023learning}). 
    The cutoff $r$, the number of layers $n_l$, the order of irreducible representations (rotation order) $\ell$, and body order $\nu$ are indicated for each MLIP. Results are reproduced from Ref.~\cite{Kim2025Universalb}.
    \figLabelCapt{b}: 
    Parity plots comparing the diagonal components of the BEC tensor ($Z^{*}_{\alpha\alpha}$) of CACE LES with RPBE-D3 reference values for 100 bulk water configurations~\cite{Schmiedmayer2024}; insets show the off-diagonal components ($Z^{*}_{\alpha\beta}$ with $\alpha\neq\beta$). 
    E+F denotes a CACE LES model trained on energy and forces, E+F+BEC is reproduced from Ref.~\cite{zhong2025machine} and was trained on a smaller set of 100 configurations with BEC labels, and E+F+Qeq incorporates a charge equilibration (Qeq) scheme~\cite{rappe1991charge, Ghasemi2015Interatomic}.
    The test RMSE values for forces in meV/\AA{} are 21.0 (E+F), 25.3 (E+F+BEC), and 21.1 (E+F+Qeq).
    All three models use the same CACE settings ($r=4.5$~\AA{}, $n_l=1$, $\nu=3$). 
    \figLabelCapt{c}: 
    The test force RMSEs and BEC $R^2$ values of MACE LES models ($r=4.5$~\AA{}, $n_l=2$, $\ell=1$) as a function of training set size.
    \figLabelCapt{d}:
    Computational performance benchmarks of MD simulations of bulk liquid water for varying system sizes ($N$) using LES-augmented MLIPs in single-precision (float32) with ASE implementations~\cite{HjorthLarsen2017atomic} performed on an NVIDIA L40S GPU (48 GB memory). The left panel shows the timing of MD simulations using the MACE models ($r=4.5$~\AA{}, $n_l=2$) with different $\ell$, both without (empty symbols and dashed lines) and with LES (filled symbols and solid lines). The right panel shows the timing of the CACE models ($r=4.5$~\AA{}, $n_l=1$, $\nu=3$), without LES (blue line), with LES (red line), and with Qeq scheme implementation (orange line). A reference line for $N^3$ scaling is included to illustrate the performance. The CACE LES model with the Qeq scheme shows cubic scaling with respect to the number of atoms.
    \figLabelCapt{e}:
    The test RMSEs for energy and force of CACE LES models ($r=4.5$~\AA{}, $n_l=1$, $\nu=2$) with varying charge output dimensions.
    \figLabelCapt{f}:
    Infrared (IR) absorption spectra of bulk liquid water obtained with the E+F, E+F+BEC, and E+F+Qeq CACE LES models. The experimental IR spectrum~\cite{Bertie1996Infrared} is included for comparison. E+F and E+F+BEC results are reproduced from Ref.~\cite{zhong2025machine}.
    }
    \label{fig:water_example}
\end{figure*}

Throughout this Perspective, for illustrative purposes, we highlight representative results on the RPBE-D3 bulk water dataset~\cite{Schmiedmayer2024}, which consists of 604 training and 50 test configurations of 64 water molecules. 
The experimental value of $\varepsilon_\infty = 1.78$ for bulk water was assumed in the BEC analysis using Eqn.~\eqref{eq:z-pbc}.

Figure~\ref{fig:water_example}a compares the test force RMSEs for the short-range baseline models (hollow bars) and their LES-augmented counterparts (solid bars) across several MLIP architectures. LES systematically reduces the force errors for all models, although the magnitude of the improvement depends on the underlying short-range architecture.

Beyond improved energies and forces, LES-enabled models recover physically meaningful electrical response properties. As shown in Ref.~\cite{Kim2025Universalb}, all LES-augmented MLIPs accurately predict BECs, indicating that the long-range electrostatics are captured in a physically consistent manner. 
As a representative example, Fig.~\ref{fig:water_example}b shows the close agreement between the DFT BECs and those predicted by the CACE LES model trained only on energy and forces (E+F).
Importantly, the inference of both long-range electrostatics and BECs within LES is also highly data efficient. As shown in Fig.~\ref{fig:water_example}c, the learning curves for the MACE LES model trained only on the energy and forces of different numbers of configurations demonstrate that accurate BEC predictions emerge rapidly and converge even faster than the forces.
The BECs are not only sensitive indicators of the correct electrostatics, but are also useful for computing electrical response properties, such as the water IR spectra shown in Fig.~\ref{fig:water_example}f.

Finally, Fig.~\ref{fig:water_example}d compares the inference speed of the short-range baselines and the LES-augmented models for MACE and CACE. The results show that the additional computational cost associated with the long-range correction is small relative to the baseline evaluation cost, confirming that LES provides an efficient route to incorporating long-range electrostatics without sacrificing performance or system size scalability.

The performance of LES, achieved with a simple algorithm and implementation, is surprising to us: long-range electrostatics for MLIPs turns out to be easier than we thought. 
In this Perspective, we distill the underlying design principles that make LES work, discuss other possible variants and extensions that also satisfy these principles, highlight the limitations and open questions, and outline our outlook for the future employment of long-range MLIPs.

\section{Two simple principles}

We argue that two simple design principles, outlined below, are sufficient for the easy and reliable incorporation of long-range electrostatics into MLIPs under the LES framework.

\textbf{Principle 1: Coulomb functional form with environment-dependent charges.}
This principle ensures that the learned long-range interaction is physically grounded: 
the one-dimensional environment-dependent charge on each atom is physically interpretable, and
the Coulomb form (Eqns.~\ref{eq:e_lr-finite} and ~\ref{eq:e_lr}) guarantees the correct asymptotic behavior and may help extrapolate beyond trained charge separation distances.

Several long-range models shown in the blue region of Fig.~\ref{fig:lr_map}, including 4G-HDNNP~\cite{ko2021fourth}, PhysNet~\cite{unke2019physnet}, BAMBOO~\cite{gong_predictive_2025}, and AIMNET2~\cite{Anstine2025}, also satisfy this principle. 
In contrast, models in the purple region of Fig.~\ref{fig:lr_map} do not explicitly rely on physical charges. 
For example, although LODE~\cite{grisafi2019incorporating, huguenin2023physics} encodes Coulomb and other asymptotically decaying potentials ($1/r^p$) in the atomic environment, its high-dimensional features cannot be directly interpreted as physical charges. Similarly, Ewald MP~\cite{kosmala2023ewald} learns long-range descriptors in reciprocal space using trainable frequency filters, without explicitly enforcing a $1/r$ interaction form.

It is difficult to assert which strategy will yield better training accuracy: physically constrained electrostatics or fully learned long-range features. 
In principle, more flexible learned formulations can capture a broader class of long-range interactions, including dispersion~\cite{huguenin2023physics, kosmala2023ewald}, and may therefore achieve lower fitting errors.
However, Coulomb forces represent the dominant contribution among long-range interactions, while most others are much weaker. 
At the same time, all long-range interactions are intrinsically weak relative to short-range forces yet involve many more atoms, making them fundamentally harder to learn from finite training datasets without inductive bias.
The empirical comparisons of these models on specific datasets are scarce in the literature. 
Fuchs et al.~\cite{fuchs2025Learninga} reported that an Allegro~\cite{musaelian2023learning} model with a charge equilibration layer achieves performance comparable to a PaiNN model~\cite{painn} with Ewald MP~\cite{kosmala2023ewald} on the OE62 dataset~\cite{Stuke2020Atomic}.
Several models, including 4G-HDNNP~\cite{ko2021fourth}, charge-constrained ACE~\cite{Rinaldi2025Chargeconstrained}, LES-augmented models~\cite{King2025Machine,Kim2025Universalb}, LOREM~\cite{Rumiantsev2025Learning}, CELLI ~\cite{fuchs2025Learninga} that uses a charge equilibration layer,
charge-equilibrated TensorNet
(QET)~\cite{Ko2025Fast}, and NequIP-LR~\cite{Maruf2025Learninga}, have been benchmarked on some of the datasets (C$_{10}$H$_{2}$/C$_{10}$H$_{3}^{+}$, Ag$^{+/-}_{3}$, Na$_{8/9}$Cl$_{8}^{+}$, and Au$_{2}$ on MgO(001) compiled by Ko et al.~\cite{ko2021fourth}. Among these, LES and LOREM have some of the lowest test energy and force errors. 
However, it has been questioned whether these datasets cover very challenging long-range interactions~\cite{Rumiantsev2025Learning}. 
Moreover, these existing comparisons often use different short-range MLIP baselines, making it unclear whether the difference in accuracy stems from the baseline itself or the long-range correction.

Here, we test how the dimensionality of the long-range features affects learning accuracy. 
In the original formulation of LES~\cite{Cheng2025Latent}, the latent charge $q^{\mathrm{les}}$ can, in principle, be multi-dimensional, with the total long-range energy obtained by aggregating the contributions from different dimensions after the Ewald summation. 
Conceptually, such multi-dimensional latent charges are analogous to the high-dimensional long-range features used in Ewald MP~\cite{kosmala2023ewald} and LODE~\cite{grisafi2019incorporating, huguenin2023physics}.
As shown in Fig.~\ref{fig:water_example}e, CACE LES models trained with more latent charge output dimensions exhibit essentially negligible and inconsistent improvements in accuracy. 
While there are alternative ways to exploit high-dimensional long-range information, such as equivariant features~\cite{Rumiantsev2025Learning}, this example demonstrates that increasing the dimensionality of long-range features does not necessarily lead to higher training accuracy.

In our view, the greatest advantage of adhering to an explicit Coulomb interaction form is interpretability. 
This interpretability extends beyond the conceptual meaning of ``charges'' and directly enables the computation of physically measurable quantities. 
As discussed in Refs.~\cite{King2025Machine, zhong2025machine, Kim2025Universalb}, the learned latent charges are predictive of molecular dipoles and higher-order multipoles and also enable the modeling of the electrical response properties such as IR spectra, ionic electrical conductivities, and ferroelectricity.
Moreover, the charges can be useful in hybrid schemes such as coupling quantum mechanical calculations with MLIPs (QM/ML)~\cite{Semelak2025Advancing} or combining MLIPs with classical force fields with fixed charges (ML/MM)~\cite{morado2025enhancing}.

\textbf{Principle 2: Not explicitly train on DFT charges.}
There is no unique way to map the continuous DFT electron density onto discrete partial charges: different partitioning schemes~\cite{sifain2018discovering, marenich2012charge, wiberg1993comparison}, such as Mulliken~\cite{mulliken1955electronic}, Hirshfeld~\cite{hirshfeld1977bonded}, and MBIS~\cite{verstraelen2016minimal}, often yield substantially different results. 
Consequently, training on one specific definition of DFT charges would imply a different long-range electrostatic description than training on another. 
Besides the uniqueness issue, a more subtle point is that the Coulomb interactions between atomic charges are screened by the rapidly responding background electrons with the relative permittivity $\varepsilon_\infty$~\cite{leontyev2003continuum,jorge2024theoretically}.
This screening is naturally absorbed in the LES formulation through learning from long-range electrostatic forces~\cite{zhong2025machine}, but would generally be absent if atomic charges were learned directly.
Consistent with this view, 
explicit training on DFT charges has also been shown to induce pathological behavior, leading to energy and force errors significantly larger than those of the short-range baseline~\cite{ko2021fourth}, such that an additional charge equilibration scheme was required to mitigate such issues~\cite{ko2021fourth}.

Avoiding training on DFT charges also offers a practical advantage. 
Most MLIP training datasets, especially those originally designed for short-range models, provide only atomic positions, energies, forces, and sometimes stress tensors. 
By not requiring charge labels, long-range augmentation becomes immediately applicable to a much broader range of existing datasets and systems.
It is worth mentioning that the LES architecture allows training on dipoles and/or BECs, which are physical quantities that can help improve the training.

Moreover, explicit training on DFT charges does not appear to improve energy and force accuracy compared to LES. For the dataset of a diatomic gold cluster Au$_2$ supported on the MgO(001) surface~\cite{ko2021fourth}, which includes both wetting and non-wetting adsorption geometries and where aluminum dopants deep beneath the surface strongly affect the electronic structure and relative stability, 
a CACE~\cite{cheng2024cartesian} model augmented with LES and trained only on energies and forces~\cite{King2025Machine} achieves nearly an order-of-magnitude lower errors than both the charge-constrained ACE model~\cite{drautz2019atomic, Rinaldi2025Chargeconstrained} and the 4G-HDNNP~\cite{ko2021fourth}.

More importantly, we argue that the LES charges learned just from energies and forces are physically meaningful: they correspond to effective partial charges governing electrostatic interactions and are predictive of measurable physical observables. 
As shown in Refs.~\cite{King2025Machine, Kim2025Universalb}, the learned latent charges yield accurate molecular dipoles and multipoles, and can further derive BECs using Eqns.~\eqref{eq:z-finite} and \eqref{eq:z-pbc}~\cite{zhong2025machine}.

For the water example, Fig.~\ref{fig:water_example}b demonstrates that the LES BECs trained on energy and forces (E+F) are in excellent agreement with the DFT reference. 
These BECs enable the calculation of electrical response properties such as IR spectra and electrical conductivities~\cite{zhong2025machine}. 
Fig.~\ref{fig:water_example}f shows the corresponding IR for bulk water: both the spectral shape and intensities are in good agreement with experiment, capturing distinct high-frequency intramolecular and low-frequency intermolecular modes.

Finally, with the predicted BECs, one can compute the electrostatic forces on atoms under an applied external electric field and perform molecular dynamics simulations under finite fields using MLIPs. 
Using this capability, we have shown that LES-augmented MLIPs can reproduce electric-field-dependent phenomena, including the IR spectral shift of bulk water under static fields, the ionic conductivity of a superionic material, and the polarization–electric field hysteresis loop of a ferroelectric material~\cite{zhong2025machine}.

\section{Other possible extensions}

Once the two design principles are met, LES can be combined with various additional strategies to suit specific applications. Here we present several examples.

\textbf{Universal compatibility with any short-range MLIPs.}
LES can serve as a universal augmentation to any short-range MLIP~\cite{Kim2025Universalb}, and we have demonstrated its integration with methods such as MACE~\cite{batatia2022mace}, NequIP~\cite{batzner20223}, Allegro~\cite{musaelian2023learning}, CACE~\cite{cheng2024cartesian}, CHGNet~\cite{deng2023chgnet}, and UMA~\cite{wood2025umafamilyuniversalmodels}. 
Among these,  MACE, NequIP, CACE, CHGNet, and UMA use node features to predict local atomic energy contributions, and the same features are straightforwardly used to predict the latent charges.
Instead, Allegro uses edge features to predict pairwise energies, which are then summed to obtain per-atom energies; in the same spirit, the LES augmentation with Allegro predicts the pairwise charge contributions first using the same edge features and then aggregates them to form the per-atom latent charges.

\textbf{Optional fine-tuning using dipoles or BECs.}
Although energies and forces alone are sufficient to recover the correct electrostatics, additional training labels, such as dipoles or BEC tensors, can be incorporated for fine-tuning. 
For example, we show that fine-tuning CACE LES models using DFT-computed BECs further improves BEC prediction accuracy while preserving energy and force accuracy for both bulk water and PbTiO$_3$ datasets~\cite{zhong2025machine}. 
The water example shown in the middle panel (E+F+BEC) of Fig.~\ref{fig:water_example}b illustrates this modest improvement in BEC predictions. 
As expected, the corresponding IR spectrum in Fig.~\ref{fig:water_example}f remains in good agreement with experiment.

\textbf{Integrating optional charge equilibration (Qeq).}
Empirically, we have found charge equilibration schemes to be unnecessary within the LES framework~\cite{Cheng2025Latent, King2025Machine}. 
For neutral systems, the sum of the learned charges is typically very close to zero; any small residual is treated as a uniform background charge, corresponding to the tinfoil boundary condition in the reciprocal-space evaluation of the electrostatic interactions in Eqn.~\eqref{eq:e_lr}. 

Nevertheless, LES can be straightforwardly combined with charge equilibration using the Qeq approach~\cite{rappe1991charge, Ghasemi2015Interatomic}, which determines the atomic charges $q_i$ by minimizing the energy
\begin{equation}
E_{\text{Qeq}} = E_{\text{elec}} + \sum_{i=1}^{N} \left( \chi_{i} q_{i} + \frac{1}{2} J_{i} q_{i}^{2} \right),
\label{eq:E_Qeq}
\end{equation}
where $E_{\text{elec}}$ is the Coulomb interaction between charges, $\chi_i$ are the atomic electronegativities, and $J_i$ are element-specific hardness parameters. 
The $\chi_i$ are predicted by a neural network from local invariant features, while the $J_i$ are assigned based on element type. 
Consistent with the LES design principles, no DFT charges are used for training.

We implemented LES combined with Qeq (LES-Qeq in Fig.~\ref{fig:lr_map}),
with trainable $\chi_i$ and $J_i$.
For the water example, the right panel (E+F+Qeq) of Fig.~\ref{fig:water_example}b shows that the addition of Qeq preserves the accurate predictions of forces and LES BECs.
The corresponding predicted IR spectrum in Fig.~\ref{fig:water_example}f also exhibits good agreement with the reference. 
However, the computational cost of Qeq is substantial: the CACE LES-Qeq model in the right panel of Fig.~\ref{fig:water_example}d shows that incorporating Qeq leads to cubic scaling due to the matrix inversion operation involved. 
As such, methods have recently been proposed to reduce this cost, including using the iterative solution of Qeq instead of direct matrix inversion~\cite{Kocer2025Iterative}, and developing a particle mesh iterative solver~\cite{Gubler2024Acceleratinga}.

\textbf{Coupling with the Siepmann–Sprik metal model.}
As illustrated by the LES-MetalWall in Fig.~\ref{fig:lr_map}, the LES can be coupled with the Siepmann–Sprik metal model~\cite{siepmann1995influence}: LES determines the charges on the electrolyte, and the Siepmann–Sprik method describes the response charges on metal electrodes. This hybrid long-range framework enables nanosecond finite-field MD simulations of electrode–electrolyte interfaces under periodic boundary conditions~\cite{Wang2025Ionmodulateda}.

\textbf{Charge and spin embedding.}
For datasets containing multiple charge or spin states, a degeneracy problem can arise: configurations that are geometrically identical but electronically distinct share the same representations in standard MLIPs, which can lead to poor model performance~\cite{ko2021fourth, Rinaldi2025Chargeconstrained}.
To lift the degeneracy, one can embed the charge and/or spin states directly in the local atomic environment features.
This can be achieved either by encoding global charge or spin states to inform the local features of all atoms~\cite{Simeon2025Broadening, King2025Machine},
or by utilizing atomwise states such as the oxidation state or the magnetic moment for each atom~\cite{Simeon2025Broadening}.
As such, not only will the predicted atomic energies depend on these states, but also the latent charges.

\textbf{Learning atomic dipoles or dispersion coefficients.}
The expression for the long-range energy in Eqns.~\eqref{eq:e_lr-finite} and \eqref{eq:e_lr} can be straightforwardly generalized to other $1/r^p$ interactions, such as charge–dipole ($p=2$), dipole–dipole ($p=3$), or dispersion-like ($p=6$) terms. 
In these cases, the learned latent variables would no longer be scalar charges but higher-order quantities: for example, equivariant atomic dipole vectors for dipolar interactions, or per-atom dispersion coefficients for van der Waals–type contributions.

\section{Limitations and open questions}
There are known limitations of the LES algorithm:
First, the latent atomic charges $q^\mathrm{les}$ are determined from local features, and thus do not provide an explicit mechanism for truly long-range charge transfer mediated by mobile carriers. 
In materials with metallic or highly delocalized electronic character, an applied external field can induce charge rearrangements that are, in principle, global. 
An example here is the induced surface charge on a macroscopic metal electrode. 
In such cases, a purely local LES scheme is not sufficient to describe global charge equilibration. 
A possible remedy is to couple LES to an explicit metallic boundary model, as in LES–MetalWall, where LES controls the electrolyte charges and the Siepmann–Sprik model performs charge equilibration on the metal electrodes~\cite{Wang2025Ionmodulateda}.

Second, extracting BEC tensors for periodic systems via Eqn.~\eqref{eq:z-pbc} requires specifying a constant high-frequency dielectric permittivity $\varepsilon_\infty$ for a homogeneous bulk material~\cite{zhong2025machine}. 
For heterogeneous systems, such as interfaces between materials with different $\varepsilon_\infty$, it is not obvious how to choose or generalize this parameter.
Developing consistent strategies to extend LES-based BEC extraction and response-property calculations to inhomogeneous or interfacial systems remains an open challenge.

Third, scalability to very large system sizes remains an engineering challenge.
The nominal $\mathcal{O}(N^{3/2})$ scaling of Ewald summation itself is not the main bottleneck, as this operation is typically inexpensive relative to other components of modern MLIPs, and more favorable scaling variants such as particle–particle particle–mesh (P3M) or particle–mesh Ewald (PME) are readily available~\cite{Loche2025Fastb}.
Rather, the main difficulty arises because interatomic forces are obtained via the automatic differentiation of the total energy with respect to atomic positions, where the total energy is a global quantity that includes the long-range electrostatic contribution, which makes it nontrivial to distribute force evaluation across multiple GPUs.

Finally, there is currently no ``foundation model''~\cite{yuan2025foundation} that spans broad regions of chemical and materials space while incorporating LES. 
Encouragingly, MACELES-OFF~\cite{Kim2025Universalb}, trained on the SPICE dataset~\cite{eastman2023spice} of molecules and small clusters, is more accurate than its short-range counterpart MACE-OFF~\cite{kovacs2025MACEOFF}, predicts dipoles and BECs reliably, and provides improved descriptions of bulk liquids. 
However, it remains an open question whether LES augmentation will offer similar advantages for substantially larger, more diverse molecular datasets or bulk materials datasets~\cite{barroso_omat24, levine2025openmolecules2025omol25}. 
Exploring this direction is key to determining whether long-range electrostatics can be made standard in next-generation foundation MLIPs.

\section{Conclusion and outlook}

In this Perspective, we examine the connections and distinctions between LES and other representative methods for learning long-range interactions (Fig.~\ref{fig:lr_map}). 
While many of the aforementioned approaches show great promise, we argue that a minimal and physically grounded design is sufficient for incorporating long-range electrostatics, based on two simple principles: (i) use a physical functional form with flexible, environment-dependent atomic charges to capture electrostatic interactions, and (ii) avoid explicit training on DFT partial charges. 
Besides, there are many avenues for other optional extensions.

Despite the known limitations, many aspects of the LES method have turned out to be easier than we initially anticipated: the design principles and algorithm are straightforward; 
only standard energy and force labels are required, 
training is data-efficient (Fig.~\ref{fig:water_example}c); 
electrical response properties can be extracted reliably (Figs.~\ref{fig:water_example}b and f); 
and the associated computational overhead is low (Fig.~\ref{fig:water_example}d).

This naturally raises a broader question: why do short-range MLIPs remain the dominant paradigm, while long-range electrostatic MLIPs have been adopted only sparingly in practical simulations? 
Equivalently, if long-range electrostatics can in fact be incorporated so simply and efficiently, why was this not widely recognized earlier? 
One plausible reason is that some of the early approaches shown in Fig.~\ref{fig:lr_map} do not fully satisfy the two design principles outlined above, or may have higher computational expense. 
Another explanation may be that long-range methods have not yet been pushed and benchmarked extensively enough to convincingly demonstrate their broad applicability and computational efficiency. 

Another possibility is simply that the atomistic modeling community has long been used to short-range MLIPs and has not yet made the transition to long-range models. 
Speaking from our own experience, when one of us (B.C.) began using MLIPs around 2018, it was already striking that atomic interactions could be learned and converged simulations performed with near–quantum mechanical accuracy. 
At that time, the systems we studied were predominantly homogeneous bulk systems, such as bulk water~\cite{cheng2019ab}, superionic water~\cite{cheng2021phase}, and high-pressure liquid hydrogen~\cite{cheng2020evidence}.  
For these systems, long-range electrostatics is conceptually important, but was not the main concern: long-range interactions are largely isotropic and tend to average out, allowing short-range models to effectively learn the mean-field forces. 
Many observables of interest, such as radial distribution functions, transport coefficients, and free energies, are not sensitive to long-range electrostatics~\cite{yue2021short}.
At the same time, the training errors of the first-generation MLIPs~\cite{behler2007generalized,bartok2010gaussian} were substantially larger than those of atomic cluster expansion (ACE)~\cite{drautz2019atomic} or MPNNs~\cite{Batatia2025designa}, due to the lack of higher-body-order terms, making long-range errors relatively less important. 
It is only more recently, as our work has shifted toward heterogeneous and interfacial systems, that long-range electrostatics has become essential.
Interfaces introduce inhomogeneity, as well as intrinsic polarization and electric fields, where the short-range MLIPs struggle.
As the community increasingly studies complex, multi-component, and field-driven systems, especially electrochemical interfaces, polar and ionic materials, and biomolecules, we expect long-range MLIPs to become not just beneficial but necessary.

\textbf{Data availability}~
The RPBE-D3 bulk water dataset, training scripts, evaluation scripts, the trained CACE E+F+Qeq model, and CACE LES and MACE LES models used to produce results in Figs.~\ref{fig:water_example}c, d, and e are available at \url{https://github.com/ChengUCB/les_fit}.

\textbf{Code availability}~
The LES library is publicly available at 
\url{https://github.com/ChengUCB/les}.
The CACE package with the LES implementation is available at \url{https://github.com/BingqingCheng/cace}.
The MACE package with the LES implementation is available at \url{https://github.com/ACEsuit/mace}.
The NequIP and Allegro LES extension package is available at \url{https://github.com/ChengUCB/NequIP-LES}.
The MatGL package with the LES implementation is available at \url{https://github.com/ChengUCB/matgl}.
The UMA package with the LES implementation is available at \url{https://github.com/santi921/fairchem/tree/les_branch}.

\textbf{Acknowledgements}~
B.C. gratefully thanks Christoph Dellago for his mentorship and influence. 
Besides his seminal contributions to statistical mechanics, Christoph Dellago is an early developer and adopter of machine learning interatomic potentials. 
B.C. did two exchanges in the groups of Christoph Dellago and Jörg Behler in 2018, with transformative impact on her research directions.\\
We thank Peichen Zhong and Daniel S. King for useful feedback on the manuscript, and for the collaborations on the LES method. \\
Funding acknowledgement: Research reported in this publication was supported by the National Institute Of General Medical Sciences of the National Institutes of Health under Award Number R35GM159986.
The content is solely the responsibility of the authors and does not necessarily represent the official views of the National Institutes of Health.

\textbf{Competing Interests}
B.C. has an equity stake in AIMATX Inc.
The University of California, Berkeley has filed a provisional patent for the Latent Ewald Summation algorithm. 



\begin{thebibliography}{101}%
\makeatletter
\providecommand \@ifxundefined [1]{%
 \@ifx{#1\undefined}
}%
\providecommand \@ifnum [1]{%
 \ifnum #1\expandafter \@firstoftwo
 \else \expandafter \@secondoftwo
 \fi
}%
\providecommand \@ifx [1]{%
 \ifx #1\expandafter \@firstoftwo
 \else \expandafter \@secondoftwo
 \fi
}%
\providecommand \natexlab [1]{#1}%
\providecommand \enquote  [1]{``#1''}%
\providecommand \bibnamefont  [1]{#1}%
\providecommand \bibfnamefont [1]{#1}%
\providecommand \citenamefont [1]{#1}%
\providecommand \href@noop [0]{\@secondoftwo}%
\providecommand \href [0]{\begingroup \@sanitize@url \@href}%
\providecommand \@href[1]{\@@startlink{#1}\@@href}%
\providecommand \@@href[1]{\endgroup#1\@@endlink}%
\providecommand \@sanitize@url [0]{\catcode `\\12\catcode `\$12\catcode `\&12\catcode `\#12\catcode `\^12\catcode `\_12\catcode `\%12\relax}%
\providecommand \@@startlink[1]{}%
\providecommand \@@endlink[0]{}%
\providecommand \url  [0]{\begingroup\@sanitize@url \@url }%
\providecommand \@url [1]{\endgroup\@href {#1}{\urlprefix }}%
\providecommand \urlprefix  [0]{URL }%
\providecommand \Eprint [0]{\href }%
\providecommand \doibase [0]{http://dx.doi.org/}%
\providecommand \selectlanguage [0]{\@gobble}%
\providecommand \bibinfo  [0]{\@secondoftwo}%
\providecommand \bibfield  [0]{\@secondoftwo}%
\providecommand \translation [1]{[#1]}%
\providecommand \BibitemOpen [0]{}%
\providecommand \bibitemStop [0]{}%
\providecommand \bibitemNoStop [0]{.\EOS\space}%
\providecommand \EOS [0]{\spacefactor3000\relax}%
\providecommand \BibitemShut  [1]{\csname bibitem#1\endcsname}%
\let\auto@bib@innerbib\@empty
\bibitem [{\citenamefont {French}\ \emph {et~al.}(2010)\citenamefont {French}, \citenamefont {Parsegian}, \citenamefont {Podgornik}, \citenamefont {Rajter}, \citenamefont {Jagota}, \citenamefont {Luo}, \citenamefont {Asthagiri}, \citenamefont {Chaudhury}, \citenamefont {Chiang}, \citenamefont {Granick} \emph {et~al.}}]{french2010long}%
  \BibitemOpen
  \bibfield  {author} {\bibinfo {author} {\bibfnamefont {Roger~H}\ \bibnamefont {French}}, \bibinfo {author} {\bibfnamefont {V~Adrian}\ \bibnamefont {Parsegian}}, \bibinfo {author} {\bibfnamefont {Rudolf}\ \bibnamefont {Podgornik}}, \bibinfo {author} {\bibfnamefont {Rick~F}\ \bibnamefont {Rajter}}, \bibinfo {author} {\bibfnamefont {Anand}\ \bibnamefont {Jagota}}, \bibinfo {author} {\bibfnamefont {Jian}\ \bibnamefont {Luo}}, \bibinfo {author} {\bibfnamefont {Dilip}\ \bibnamefont {Asthagiri}}, \bibinfo {author} {\bibfnamefont {Manoj~K}\ \bibnamefont {Chaudhury}}, \bibinfo {author} {\bibfnamefont {Yet-ming}\ \bibnamefont {Chiang}}, \bibinfo {author} {\bibfnamefont {Steve}\ \bibnamefont {Granick}},  \emph {et~al.},\ }\bibfield  {title} {\enquote {\bibinfo {title} {Long range interactions in nanoscale science},}\ }\href@noop {} {\bibfield  {journal} {\bibinfo  {journal} {Reviews of Modern Physics}\ }\textbf {\bibinfo {volume} {82}},\ \bibinfo {pages} {1887--1944} (\bibinfo {year} {2010})}\BibitemShut {NoStop}%
\bibitem [{\citenamefont {York}\ \emph {et~al.}(1995)\citenamefont {York}, \citenamefont {Yang}, \citenamefont {Lee}, \citenamefont {Darden},\ and\ \citenamefont {Pedersen}}]{York1995Accurate}%
  \BibitemOpen
  \bibfield  {author} {\bibinfo {author} {\bibfnamefont {Darrin~M.}\ \bibnamefont {York}}, \bibinfo {author} {\bibfnamefont {Weitao}\ \bibnamefont {Yang}}, \bibinfo {author} {\bibfnamefont {Hsing}\ \bibnamefont {Lee}}, \bibinfo {author} {\bibfnamefont {Tom}\ \bibnamefont {Darden}}, \ and\ \bibinfo {author} {\bibfnamefont {Lee~G.}\ \bibnamefont {Pedersen}},\ }\bibfield  {title} {\enquote {\bibinfo {title} {Toward the accurate modeling of dna: The importance of long-range electrostatics},}\ }\href {\doibase 10.1021/ja00122a034} {\bibfield  {journal} {\bibinfo  {journal} {Journal of the American Chemical Society}\ }\textbf {\bibinfo {volume} {117}},\ \bibinfo {pages} {5001--5002} (\bibinfo {year} {1995})}\BibitemShut {NoStop}%
\bibitem [{\citenamefont {Rodgers}\ and\ \citenamefont {Weeks}(2008)}]{rodgers2008interplay}%
  \BibitemOpen
  \bibfield  {author} {\bibinfo {author} {\bibfnamefont {Jocelyn~M}\ \bibnamefont {Rodgers}}\ and\ \bibinfo {author} {\bibfnamefont {John~D}\ \bibnamefont {Weeks}},\ }\bibfield  {title} {\enquote {\bibinfo {title} {Interplay of local hydrogen-bonding and long-ranged dipolar forces in simulations of confined water},}\ }\href@noop {} {\bibfield  {journal} {\bibinfo  {journal} {Proceedings of the National Academy of Sciences}\ }\textbf {\bibinfo {volume} {105}},\ \bibinfo {pages} {19136--19141} (\bibinfo {year} {2008})}\BibitemShut {NoStop}%
\bibitem [{\citenamefont {Prodan}\ and\ \citenamefont {Kohn}(2005)}]{prodan2005nearsightedness}%
  \BibitemOpen
  \bibfield  {author} {\bibinfo {author} {\bibfnamefont {Emil}\ \bibnamefont {Prodan}}\ and\ \bibinfo {author} {\bibfnamefont {Walter}\ \bibnamefont {Kohn}},\ }\bibfield  {title} {\enquote {\bibinfo {title} {Nearsightedness of electronic matter},}\ }\href@noop {} {\bibfield  {journal} {\bibinfo  {journal} {Proceedings of the National Academy of Sciences}\ }\textbf {\bibinfo {volume} {102}},\ \bibinfo {pages} {11635--11638} (\bibinfo {year} {2005})}\BibitemShut {NoStop}%
\bibitem [{\citenamefont {Keith}\ \emph {et~al.}(2021)\citenamefont {Keith}, \citenamefont {Vassilev-Galindo}, \citenamefont {Cheng}, \citenamefont {Chmiela}, \citenamefont {Gastegger}, \citenamefont {M{\"u}ller},\ and\ \citenamefont {Tkatchenko}}]{keith2021combining}%
  \BibitemOpen
  \bibfield  {author} {\bibinfo {author} {\bibfnamefont {John~A}\ \bibnamefont {Keith}}, \bibinfo {author} {\bibfnamefont {Valentin}\ \bibnamefont {Vassilev-Galindo}}, \bibinfo {author} {\bibfnamefont {Bingqing}\ \bibnamefont {Cheng}}, \bibinfo {author} {\bibfnamefont {Stefan}\ \bibnamefont {Chmiela}}, \bibinfo {author} {\bibfnamefont {Michael}\ \bibnamefont {Gastegger}}, \bibinfo {author} {\bibfnamefont {Klaus-Robert}\ \bibnamefont {M{\"u}ller}}, \ and\ \bibinfo {author} {\bibfnamefont {Alexandre}\ \bibnamefont {Tkatchenko}},\ }\bibfield  {title} {\enquote {\bibinfo {title} {Combining machine learning and computational chemistry for predictive insights into chemical systems},}\ }\href@noop {} {\bibfield  {journal} {\bibinfo  {journal} {Chemical reviews}\ }\textbf {\bibinfo {volume} {121}},\ \bibinfo {pages} {9816--9872} (\bibinfo {year} {2021})}\BibitemShut {NoStop}%
\bibitem [{\citenamefont {Unke}\ \emph {et~al.}(2021{\natexlab{a}})\citenamefont {Unke}, \citenamefont {Chmiela}, \citenamefont {Sauceda}, \citenamefont {Gastegger}, \citenamefont {Poltavsky}, \citenamefont {Sch{\"u}tt}, \citenamefont {Tkatchenko},\ and\ \citenamefont {M{\"u}ller}}]{unke2021machine}%
  \BibitemOpen
  \bibfield  {author} {\bibinfo {author} {\bibfnamefont {Oliver~T}\ \bibnamefont {Unke}}, \bibinfo {author} {\bibfnamefont {Stefan}\ \bibnamefont {Chmiela}}, \bibinfo {author} {\bibfnamefont {Huziel~E}\ \bibnamefont {Sauceda}}, \bibinfo {author} {\bibfnamefont {Michael}\ \bibnamefont {Gastegger}}, \bibinfo {author} {\bibfnamefont {Igor}\ \bibnamefont {Poltavsky}}, \bibinfo {author} {\bibfnamefont {Kristof~T}\ \bibnamefont {Sch{\"u}tt}}, \bibinfo {author} {\bibfnamefont {Alexandre}\ \bibnamefont {Tkatchenko}}, \ and\ \bibinfo {author} {\bibfnamefont {Klaus-Robert}\ \bibnamefont {M{\"u}ller}},\ }\bibfield  {title} {\enquote {\bibinfo {title} {Machine learning force fields},}\ }\href@noop {} {\bibfield  {journal} {\bibinfo  {journal} {Chemical Reviews}\ }\textbf {\bibinfo {volume} {121}},\ \bibinfo {pages} {10142--10186} (\bibinfo {year} {2021}{\natexlab{a}})}\BibitemShut {NoStop}%
\bibitem [{\citenamefont {Behler}\ and\ \citenamefont {Parrinello}(2007)}]{behler2007generalized}%
  \BibitemOpen
  \bibfield  {author} {\bibinfo {author} {\bibfnamefont {J{\"o}rg}\ \bibnamefont {Behler}}\ and\ \bibinfo {author} {\bibfnamefont {Michele}\ \bibnamefont {Parrinello}},\ }\bibfield  {title} {\enquote {\bibinfo {title} {Generalized neural-network representation of high-dimensional potential-energy surfaces},}\ }\href@noop {} {\bibfield  {journal} {\bibinfo  {journal} {Physical review letters}\ }\textbf {\bibinfo {volume} {98}},\ \bibinfo {pages} {146401} (\bibinfo {year} {2007})}\BibitemShut {NoStop}%
\bibitem [{\citenamefont {Bart{\'o}k}\ \emph {et~al.}(2010)\citenamefont {Bart{\'o}k}, \citenamefont {Payne}, \citenamefont {Kondor},\ and\ \citenamefont {Cs{\'a}nyi}}]{bartok2010gaussian}%
  \BibitemOpen
  \bibfield  {author} {\bibinfo {author} {\bibfnamefont {Albert~P}\ \bibnamefont {Bart{\'o}k}}, \bibinfo {author} {\bibfnamefont {Mike~C}\ \bibnamefont {Payne}}, \bibinfo {author} {\bibfnamefont {Risi}\ \bibnamefont {Kondor}}, \ and\ \bibinfo {author} {\bibfnamefont {G{\'a}bor}\ \bibnamefont {Cs{\'a}nyi}},\ }\bibfield  {title} {\enquote {\bibinfo {title} {Gaussian approximation potentials: The accuracy of quantum mechanics, without the electrons},}\ }\href@noop {} {\bibfield  {journal} {\bibinfo  {journal} {Physical review letters}\ }\textbf {\bibinfo {volume} {104}},\ \bibinfo {pages} {136403} (\bibinfo {year} {2010})}\BibitemShut {NoStop}%
\bibitem [{\citenamefont {Sch{\"u}tt}\ \emph {et~al.}(2017)\citenamefont {Sch{\"u}tt}, \citenamefont {Kindermans}, \citenamefont {Sauceda~Felix}, \citenamefont {Chmiela}, \citenamefont {Tkatchenko},\ and\ \citenamefont {M{\"u}ller}}]{schutt2017schnet}%
  \BibitemOpen
  \bibfield  {author} {\bibinfo {author} {\bibfnamefont {Kristof}\ \bibnamefont {Sch{\"u}tt}}, \bibinfo {author} {\bibfnamefont {Pieter-Jan}\ \bibnamefont {Kindermans}}, \bibinfo {author} {\bibfnamefont {Huziel~Enoc}\ \bibnamefont {Sauceda~Felix}}, \bibinfo {author} {\bibfnamefont {Stefan}\ \bibnamefont {Chmiela}}, \bibinfo {author} {\bibfnamefont {Alexandre}\ \bibnamefont {Tkatchenko}}, \ and\ \bibinfo {author} {\bibfnamefont {Klaus-Robert}\ \bibnamefont {M{\"u}ller}},\ }\bibfield  {title} {\enquote {\bibinfo {title} {Schnet: A continuous-filter convolutional neural network for modeling quantum interactions},}\ }\href@noop {} {\bibfield  {journal} {\bibinfo  {journal} {Advances in neural information processing systems}\ }\textbf {\bibinfo {volume} {30}} (\bibinfo {year} {2017})}\BibitemShut {NoStop}%
\bibitem [{\citenamefont {Haghighatlari}\ \emph {et~al.}(2022)\citenamefont {Haghighatlari}, \citenamefont {Li}, \citenamefont {Guan}, \citenamefont {Zhang}, \citenamefont {Das}, \citenamefont {J.~Stein}, \citenamefont {{Heidar-Zadeh}}, \citenamefont {Liu}, \citenamefont {{Head-Gordon}}, \citenamefont {Bertels}, \citenamefont {Hao}, \citenamefont {Leven},\ and\ \citenamefont {{Head-Gordon}}}]{haghighatlari2022NewtonNet}%
  \BibitemOpen
  \bibfield  {author} {\bibinfo {author} {\bibfnamefont {Mojtaba}\ \bibnamefont {Haghighatlari}}, \bibinfo {author} {\bibfnamefont {Jie}\ \bibnamefont {Li}}, \bibinfo {author} {\bibfnamefont {Xingyi}\ \bibnamefont {Guan}}, \bibinfo {author} {\bibfnamefont {Oufan}\ \bibnamefont {Zhang}}, \bibinfo {author} {\bibfnamefont {Akshaya}\ \bibnamefont {Das}}, \bibinfo {author} {\bibfnamefont {Christopher}\ \bibnamefont {J.~Stein}}, \bibinfo {author} {\bibfnamefont {Farnaz}\ \bibnamefont {{Heidar-Zadeh}}}, \bibinfo {author} {\bibfnamefont {Meili}\ \bibnamefont {Liu}}, \bibinfo {author} {\bibfnamefont {Martin}\ \bibnamefont {{Head-Gordon}}}, \bibinfo {author} {\bibfnamefont {Luke}\ \bibnamefont {Bertels}}, \bibinfo {author} {\bibfnamefont {Hongxia}\ \bibnamefont {Hao}}, \bibinfo {author} {\bibfnamefont {Itai}\ \bibnamefont {Leven}}, \ and\ \bibinfo {author} {\bibfnamefont {Teresa}\ \bibnamefont {{Head-Gordon}}},\ }\bibfield  {title} {\enquote {\bibinfo {title} {Newtonnet: A newtonian message passing network for deep
  learning of interatomic potentials and forces},}\ }\href {\doibase 10.1039/D2DD00008C} {\bibfield  {journal} {\bibinfo  {journal} {Digital Discovery}\ }\textbf {\bibinfo {volume} {1}},\ \bibinfo {pages} {333--343} (\bibinfo {year} {2022})}\BibitemShut {NoStop}%
\bibitem [{\citenamefont {Batzner}\ \emph {et~al.}(2022)\citenamefont {Batzner}, \citenamefont {Musaelian}, \citenamefont {Sun}, \citenamefont {Geiger}, \citenamefont {Mailoa}, \citenamefont {Kornbluth}, \citenamefont {Molinari}, \citenamefont {Smidt},\ and\ \citenamefont {Kozinsky}}]{batzner20223}%
  \BibitemOpen
  \bibfield  {author} {\bibinfo {author} {\bibfnamefont {Simon}\ \bibnamefont {Batzner}}, \bibinfo {author} {\bibfnamefont {Albert}\ \bibnamefont {Musaelian}}, \bibinfo {author} {\bibfnamefont {Lixin}\ \bibnamefont {Sun}}, \bibinfo {author} {\bibfnamefont {Mario}\ \bibnamefont {Geiger}}, \bibinfo {author} {\bibfnamefont {Jonathan~P}\ \bibnamefont {Mailoa}}, \bibinfo {author} {\bibfnamefont {Mordechai}\ \bibnamefont {Kornbluth}}, \bibinfo {author} {\bibfnamefont {Nicola}\ \bibnamefont {Molinari}}, \bibinfo {author} {\bibfnamefont {Tess~E}\ \bibnamefont {Smidt}}, \ and\ \bibinfo {author} {\bibfnamefont {Boris}\ \bibnamefont {Kozinsky}},\ }\bibfield  {title} {\enquote {\bibinfo {title} {E (3)-equivariant graph neural networks for data-efficient and accurate interatomic potentials},}\ }\href@noop {} {\bibfield  {journal} {\bibinfo  {journal} {Nature communications}\ }\textbf {\bibinfo {volume} {13}},\ \bibinfo {pages} {2453} (\bibinfo {year} {2022})}\BibitemShut {NoStop}%
\bibitem [{\citenamefont {Batatia}\ \emph {et~al.}(2022)\citenamefont {Batatia}, \citenamefont {Kovacs}, \citenamefont {Simm}, \citenamefont {Ortner},\ and\ \citenamefont {Cs{\'a}nyi}}]{batatia2022mace}%
  \BibitemOpen
  \bibfield  {author} {\bibinfo {author} {\bibfnamefont {Ilyes}\ \bibnamefont {Batatia}}, \bibinfo {author} {\bibfnamefont {David~P}\ \bibnamefont {Kovacs}}, \bibinfo {author} {\bibfnamefont {Gregor}\ \bibnamefont {Simm}}, \bibinfo {author} {\bibfnamefont {Christoph}\ \bibnamefont {Ortner}}, \ and\ \bibinfo {author} {\bibfnamefont {G{\'a}bor}\ \bibnamefont {Cs{\'a}nyi}},\ }\bibfield  {title} {\enquote {\bibinfo {title} {Mace: Higher order equivariant message passing neural networks for fast and accurate force fields},}\ }\href@noop {} {\bibfield  {journal} {\bibinfo  {journal} {Advances in Neural Information Processing Systems}\ }\textbf {\bibinfo {volume} {35}},\ \bibinfo {pages} {11423--11436} (\bibinfo {year} {2022})}\BibitemShut {NoStop}%
\bibitem [{\citenamefont {Fu}\ \emph {et~al.}(2025)\citenamefont {Fu}, \citenamefont {Wood}, \citenamefont {{Barroso-Luque}}, \citenamefont {Levine}, \citenamefont {Gao}, \citenamefont {Dzamba},\ and\ \citenamefont {Zitnick}}]{fu2025Learning}%
  \BibitemOpen
  \bibfield  {author} {\bibinfo {author} {\bibfnamefont {Xiang}\ \bibnamefont {Fu}}, \bibinfo {author} {\bibfnamefont {Brandon~M.}\ \bibnamefont {Wood}}, \bibinfo {author} {\bibfnamefont {Luis}\ \bibnamefont {{Barroso-Luque}}}, \bibinfo {author} {\bibfnamefont {Daniel~S.}\ \bibnamefont {Levine}}, \bibinfo {author} {\bibfnamefont {Meng}\ \bibnamefont {Gao}}, \bibinfo {author} {\bibfnamefont {Misko}\ \bibnamefont {Dzamba}}, \ and\ \bibinfo {author} {\bibfnamefont {C.~Lawrence}\ \bibnamefont {Zitnick}},\ }\bibfield  {title} {\enquote {\bibinfo {title} {Learning smooth and expressive interatomic potentials for physical property prediction},}\ }\href@noop {} {\bibfield  {journal} {\bibinfo  {journal} {arXiv preprint arXiv:2502.12147}\ } (\bibinfo {year} {2025})}\BibitemShut {NoStop}%
\bibitem [{\citenamefont {Wood}\ \emph {et~al.}(2025)\citenamefont {Wood}, \citenamefont {Dzamba}, \citenamefont {Fu}, \citenamefont {Gao}, \citenamefont {Shuaibi}, \citenamefont {Barroso-Luque}, \citenamefont {Abdelmaqsoud}, \citenamefont {Gharakhanyan}, \citenamefont {Kitchin}, \citenamefont {Levine}, \citenamefont {Michel}, \citenamefont {Sriram}, \citenamefont {Cohen}, \citenamefont {Das}, \citenamefont {Rizvi}, \citenamefont {Sahoo}, \citenamefont {Ulissi},\ and\ \citenamefont {Zitnick}}]{wood2025umafamilyuniversalmodels}%
  \BibitemOpen
  \bibfield  {author} {\bibinfo {author} {\bibfnamefont {Brandon~M.}\ \bibnamefont {Wood}}, \bibinfo {author} {\bibfnamefont {Misko}\ \bibnamefont {Dzamba}}, \bibinfo {author} {\bibfnamefont {Xiang}\ \bibnamefont {Fu}}, \bibinfo {author} {\bibfnamefont {Meng}\ \bibnamefont {Gao}}, \bibinfo {author} {\bibfnamefont {Muhammed}\ \bibnamefont {Shuaibi}}, \bibinfo {author} {\bibfnamefont {Luis}\ \bibnamefont {Barroso-Luque}}, \bibinfo {author} {\bibfnamefont {Kareem}\ \bibnamefont {Abdelmaqsoud}}, \bibinfo {author} {\bibfnamefont {Vahe}\ \bibnamefont {Gharakhanyan}}, \bibinfo {author} {\bibfnamefont {John~R.}\ \bibnamefont {Kitchin}}, \bibinfo {author} {\bibfnamefont {Daniel~S.}\ \bibnamefont {Levine}}, \bibinfo {author} {\bibfnamefont {Kyle}\ \bibnamefont {Michel}}, \bibinfo {author} {\bibfnamefont {Anuroop}\ \bibnamefont {Sriram}}, \bibinfo {author} {\bibfnamefont {Taco}\ \bibnamefont {Cohen}}, \bibinfo {author} {\bibfnamefont {Abhishek}\ \bibnamefont {Das}}, \bibinfo {author} {\bibfnamefont {Ammar}\ \bibnamefont
  {Rizvi}}, \bibinfo {author} {\bibfnamefont {Sushree~Jagriti}\ \bibnamefont {Sahoo}}, \bibinfo {author} {\bibfnamefont {Zachary~W.}\ \bibnamefont {Ulissi}}, \ and\ \bibinfo {author} {\bibfnamefont {C.~Lawrence}\ \bibnamefont {Zitnick}},\ }\bibfield  {title} {\enquote {\bibinfo {title} {Uma: A family of universal models for atoms},}\ }\href@noop {} {\bibfield  {journal} {\bibinfo  {journal} {arXiv preprint arXiv:2506.23971}\ } (\bibinfo {year} {2025})}\BibitemShut {NoStop}%
\bibitem [{\citenamefont {Li}\ \emph {et~al.}(2018)\citenamefont {Li}, \citenamefont {Han},\ and\ \citenamefont {Wu}}]{Li2018Deeper}%
  \BibitemOpen
  \bibfield  {author} {\bibinfo {author} {\bibfnamefont {Qimai}\ \bibnamefont {Li}}, \bibinfo {author} {\bibfnamefont {Zhichao}\ \bibnamefont {Han}}, \ and\ \bibinfo {author} {\bibfnamefont {Xiao-Ming}\ \bibnamefont {Wu}},\ }\bibfield  {title} {\enquote {\bibinfo {title} {Deeper insights into graph convolutional networks for semi-supervised learning},}\ }\href@noop {} {\bibfield  {journal} {\bibinfo  {journal} {arXiv preprintarXiv:1801.07606}\ } (\bibinfo {year} {2018})}\BibitemShut {NoStop}%
\bibitem [{\citenamefont {Alon}\ and\ \citenamefont {Yahav}(2021)}]{Alon2021Bottleneck}%
  \BibitemOpen
  \bibfield  {author} {\bibinfo {author} {\bibfnamefont {Uri}\ \bibnamefont {Alon}}\ and\ \bibinfo {author} {\bibfnamefont {Eran}\ \bibnamefont {Yahav}},\ }\bibfield  {title} {\enquote {\bibinfo {title} {On the bottleneck of graph neural networks and its practical implications},}\ }\href@noop {} {\  (\bibinfo {year} {2021})}\BibitemShut {NoStop}%
\bibitem [{\citenamefont {Ko}\ \emph {et~al.}(2021{\natexlab{a}})\citenamefont {Ko}, \citenamefont {Finkler}, \citenamefont {Goedecker},\ and\ \citenamefont {Behler}}]{Ko2021GeneralPurpose}%
  \BibitemOpen
  \bibfield  {author} {\bibinfo {author} {\bibfnamefont {Tsz~Wai}\ \bibnamefont {Ko}}, \bibinfo {author} {\bibfnamefont {Jonas~A.}\ \bibnamefont {Finkler}}, \bibinfo {author} {\bibfnamefont {Stefan}\ \bibnamefont {Goedecker}}, \ and\ \bibinfo {author} {\bibfnamefont {J{\"o}rg}\ \bibnamefont {Behler}},\ }\bibfield  {title} {\enquote {\bibinfo {title} {General-purpose machine learning potentials capturing nonlocal charge transfer},}\ }\href {\doibase 10.1021/acs.accounts.0c00689} {\bibfield  {journal} {\bibinfo  {journal} {Accounts of Chemical Research}\ }\textbf {\bibinfo {volume} {54}},\ \bibinfo {pages} {808--817} (\bibinfo {year} {2021}{\natexlab{a}})}\BibitemShut {NoStop}%
\bibitem [{\citenamefont {Kocer}\ \emph {et~al.}(2022)\citenamefont {Kocer}, \citenamefont {Ko},\ and\ \citenamefont {Behler}}]{Kocer2022Neural}%
  \BibitemOpen
  \bibfield  {author} {\bibinfo {author} {\bibfnamefont {Emir}\ \bibnamefont {Kocer}}, \bibinfo {author} {\bibfnamefont {Tsz~Wai}\ \bibnamefont {Ko}}, \ and\ \bibinfo {author} {\bibfnamefont {J{\"o}rg}\ \bibnamefont {Behler}},\ }\bibfield  {title} {\enquote {\bibinfo {title} {Neural network potentials: A concise overview of methods},}\ }\href {\doibase 10.1146/annurev-physchem-082720-034254} {\bibfield  {journal} {\bibinfo  {journal} {Annual Review of Physical Chemistry}\ }\textbf {\bibinfo {volume} {73}},\ \bibinfo {pages} {163--186} (\bibinfo {year} {2022})}\BibitemShut {NoStop}%
\bibitem [{\citenamefont {Behler}(2021)}]{Behler2021Four}%
  \BibitemOpen
  \bibfield  {author} {\bibinfo {author} {\bibfnamefont {J{\"o}rg}\ \bibnamefont {Behler}},\ }\bibfield  {title} {\enquote {\bibinfo {title} {Four generations of high-dimensional neural network potentials},}\ }\href {\doibase 10.1021/acs.chemrev.0c00868} {\bibfield  {journal} {\bibinfo  {journal} {Chemical Reviews}\ }\textbf {\bibinfo {volume} {121}},\ \bibinfo {pages} {10037--10072} (\bibinfo {year} {2021})}\BibitemShut {NoStop}%
\bibitem [{\citenamefont {Anstine}\ and\ \citenamefont {Isayev}(2023)}]{Anstine2023Machinea}%
  \BibitemOpen
  \bibfield  {author} {\bibinfo {author} {\bibfnamefont {Dylan~M.}\ \bibnamefont {Anstine}}\ and\ \bibinfo {author} {\bibfnamefont {Olexandr}\ \bibnamefont {Isayev}},\ }\bibfield  {title} {\enquote {\bibinfo {title} {Machine learning interatomic potentials and long-range physics},}\ }\href {\doibase 10.1021/acs.jpca.2c06778} {\bibfield  {journal} {\bibinfo  {journal} {The Journal of Physical Chemistry A}\ }\textbf {\bibinfo {volume} {127}},\ \bibinfo {pages} {2417--2431} (\bibinfo {year} {2023})}\BibitemShut {NoStop}%
\bibitem [{\citenamefont {Cheng}(2025)}]{Cheng2025Latent}%
  \BibitemOpen
  \bibfield  {author} {\bibinfo {author} {\bibfnamefont {Bingqing}\ \bibnamefont {Cheng}},\ }\bibfield  {title} {\enquote {\bibinfo {title} {Latent ewald summation for machine learning of long-range interactions},}\ }\href {\doibase 10.1038/s41524-025-01577-7} {\bibfield  {journal} {\bibinfo  {journal} {npj Computational Materials}\ }\textbf {\bibinfo {volume} {11}},\ \bibinfo {pages} {80} (\bibinfo {year} {2025})}\BibitemShut {NoStop}%
\bibitem [{\citenamefont {Unke}\ and\ \citenamefont {Meuwly}(2019)}]{unke2019physnet}%
  \BibitemOpen
  \bibfield  {author} {\bibinfo {author} {\bibfnamefont {Oliver~T}\ \bibnamefont {Unke}}\ and\ \bibinfo {author} {\bibfnamefont {Markus}\ \bibnamefont {Meuwly}},\ }\bibfield  {title} {\enquote {\bibinfo {title} {Physnet: A neural network for predicting energies, forces, dipole moments, and partial charges},}\ }\href@noop {} {\bibfield  {journal} {\bibinfo  {journal} {Journal of chemical theory and computation}\ }\textbf {\bibinfo {volume} {15}},\ \bibinfo {pages} {3678--3693} (\bibinfo {year} {2019})}\BibitemShut {NoStop}%
\bibitem [{\citenamefont {Ko}\ \emph {et~al.}(2021{\natexlab{b}})\citenamefont {Ko}, \citenamefont {Finkler}, \citenamefont {Goedecker},\ and\ \citenamefont {Behler}}]{ko2021fourth}%
  \BibitemOpen
  \bibfield  {author} {\bibinfo {author} {\bibfnamefont {Tsz~Wai}\ \bibnamefont {Ko}}, \bibinfo {author} {\bibfnamefont {Jonas~A}\ \bibnamefont {Finkler}}, \bibinfo {author} {\bibfnamefont {Stefan}\ \bibnamefont {Goedecker}}, \ and\ \bibinfo {author} {\bibfnamefont {J{\"o}rg}\ \bibnamefont {Behler}},\ }\bibfield  {title} {\enquote {\bibinfo {title} {A fourth-generation high-dimensional neural network potential with accurate electrostatics including non-local charge transfer},}\ }\href@noop {} {\bibfield  {journal} {\bibinfo  {journal} {Nature communications}\ }\textbf {\bibinfo {volume} {12}},\ \bibinfo {pages} {398} (\bibinfo {year} {2021}{\natexlab{b}})}\BibitemShut {NoStop}%
\bibitem [{\citenamefont {Sifain}\ \emph {et~al.}(2018)\citenamefont {Sifain}, \citenamefont {Lubbers}, \citenamefont {Nebgen}, \citenamefont {Smith}, \citenamefont {Lokhov}, \citenamefont {Isayev}, \citenamefont {Roitberg}, \citenamefont {Barros},\ and\ \citenamefont {Tretiak}}]{sifain2018discovering}%
  \BibitemOpen
  \bibfield  {author} {\bibinfo {author} {\bibfnamefont {Andrew~E}\ \bibnamefont {Sifain}}, \bibinfo {author} {\bibfnamefont {Nicholas}\ \bibnamefont {Lubbers}}, \bibinfo {author} {\bibfnamefont {Benjamin~T}\ \bibnamefont {Nebgen}}, \bibinfo {author} {\bibfnamefont {Justin~S}\ \bibnamefont {Smith}}, \bibinfo {author} {\bibfnamefont {Andrey~Y}\ \bibnamefont {Lokhov}}, \bibinfo {author} {\bibfnamefont {Olexandr}\ \bibnamefont {Isayev}}, \bibinfo {author} {\bibfnamefont {Adrian~E}\ \bibnamefont {Roitberg}}, \bibinfo {author} {\bibfnamefont {Kipton}\ \bibnamefont {Barros}}, \ and\ \bibinfo {author} {\bibfnamefont {Sergei}\ \bibnamefont {Tretiak}},\ }\bibfield  {title} {\enquote {\bibinfo {title} {Discovering a transferable charge assignment model using machine learning},}\ }\href@noop {} {\bibfield  {journal} {\bibinfo  {journal} {The journal of physical chemistry letters}\ }\textbf {\bibinfo {volume} {9}},\ \bibinfo {pages} {4495--4501} (\bibinfo {year} {2018})}\BibitemShut {NoStop}%
\bibitem [{\citenamefont {Gong}\ \emph {et~al.}(2025)\citenamefont {Gong}, \citenamefont {Zhang}, \citenamefont {Mu}, \citenamefont {Pu}, \citenamefont {Wang}, \citenamefont {Han}, \citenamefont {Yu}, \citenamefont {Chen}, \citenamefont {Zheng}, \citenamefont {Wang}, \citenamefont {Chen}, \citenamefont {Yang}, \citenamefont {Wu}, \citenamefont {Shi}, \citenamefont {Gao}, \citenamefont {Yan},\ and\ \citenamefont {Xiang}}]{gong_predictive_2025}%
  \BibitemOpen
  \bibfield  {author} {\bibinfo {author} {\bibfnamefont {Sheng}\ \bibnamefont {Gong}}, \bibinfo {author} {\bibfnamefont {Yumin}\ \bibnamefont {Zhang}}, \bibinfo {author} {\bibfnamefont {Zhenliang}\ \bibnamefont {Mu}}, \bibinfo {author} {\bibfnamefont {Zhichen}\ \bibnamefont {Pu}}, \bibinfo {author} {\bibfnamefont {Hongyi}\ \bibnamefont {Wang}}, \bibinfo {author} {\bibfnamefont {Xu}~\bibnamefont {Han}}, \bibinfo {author} {\bibfnamefont {Zhiao}\ \bibnamefont {Yu}}, \bibinfo {author} {\bibfnamefont {Mengyi}\ \bibnamefont {Chen}}, \bibinfo {author} {\bibfnamefont {Tianze}\ \bibnamefont {Zheng}}, \bibinfo {author} {\bibfnamefont {Zhi}\ \bibnamefont {Wang}}, \bibinfo {author} {\bibfnamefont {Lifei}\ \bibnamefont {Chen}}, \bibinfo {author} {\bibfnamefont {Zhenze}\ \bibnamefont {Yang}}, \bibinfo {author} {\bibfnamefont {Xiaojie}\ \bibnamefont {Wu}}, \bibinfo {author} {\bibfnamefont {Shaochen}\ \bibnamefont {Shi}}, \bibinfo {author} {\bibfnamefont {Weihao}\ \bibnamefont {Gao}}, \bibinfo {author} {\bibfnamefont {Wen}\
  \bibnamefont {Yan}}, \ and\ \bibinfo {author} {\bibfnamefont {Liang}\ \bibnamefont {Xiang}},\ }\bibfield  {title} {\enquote {\bibinfo {title} {A predictive machine learning force-field framework for liquid electrolyte development},}\ }\href {\doibase 10.1038/s42256-025-01009-7} {\bibfield  {journal} {\bibinfo  {journal} {Nature Machine Intelligence}\ }\textbf {\bibinfo {volume} {7}},\ \bibinfo {pages} {543--552} (\bibinfo {year} {2025})}\BibitemShut {NoStop}%
\bibitem [{\citenamefont {Shaidu}\ \emph {et~al.}(2024)\citenamefont {Shaidu}, \citenamefont {Pellegrini}, \citenamefont {K{\"u}{\c{c}}{\"u}kbenli}, \citenamefont {Lot},\ and\ \citenamefont {de~Gironcoli}}]{shaidu2024incorporating}%
  \BibitemOpen
  \bibfield  {author} {\bibinfo {author} {\bibfnamefont {Yusuf}\ \bibnamefont {Shaidu}}, \bibinfo {author} {\bibfnamefont {Franco}\ \bibnamefont {Pellegrini}}, \bibinfo {author} {\bibfnamefont {Emine}\ \bibnamefont {K{\"u}{\c{c}}{\"u}kbenli}}, \bibinfo {author} {\bibfnamefont {Ruggero}\ \bibnamefont {Lot}}, \ and\ \bibinfo {author} {\bibfnamefont {Stefano}\ \bibnamefont {de~Gironcoli}},\ }\bibfield  {title} {\enquote {\bibinfo {title} {Incorporating long-range electrostatics in neural network potentials via variational charge equilibration from shortsighted ingredients},}\ }\href@noop {} {\bibfield  {journal} {\bibinfo  {journal} {npj Computational Materials}\ }\textbf {\bibinfo {volume} {10}},\ \bibinfo {pages} {47} (\bibinfo {year} {2024})}\BibitemShut {NoStop}%
\bibitem [{\citenamefont {Song}\ \emph {et~al.}(2024)\citenamefont {Song}, \citenamefont {Han}, \citenamefont {Henkelman},\ and\ \citenamefont {Li}}]{Song2024ChargeOptimizeda}%
  \BibitemOpen
  \bibfield  {author} {\bibinfo {author} {\bibfnamefont {Zichen}\ \bibnamefont {Song}}, \bibinfo {author} {\bibfnamefont {Jian}\ \bibnamefont {Han}}, \bibinfo {author} {\bibfnamefont {Graeme}\ \bibnamefont {Henkelman}}, \ and\ \bibinfo {author} {\bibfnamefont {Lei}\ \bibnamefont {Li}},\ }\bibfield  {title} {\enquote {\bibinfo {title} {Charge-optimized electrostatic interaction atom-centered neural network algorithm},}\ }\href {\doibase 10.1021/acs.jctc.3c01254} {\bibfield  {journal} {\bibinfo  {journal} {Journal of Chemical Theory and Computation}\ }\textbf {\bibinfo {volume} {20}},\ \bibinfo {pages} {2088--2097} (\bibinfo {year} {2024})}\BibitemShut {NoStop}%
\bibitem [{\citenamefont {Kabylda}\ \emph {et~al.}(2025)\citenamefont {Kabylda}, \citenamefont {Frank}, \citenamefont {Dou}, \citenamefont {Khabibrakhmanov}, \citenamefont {Sandonas}, \citenamefont {Unke}, \citenamefont {Chmiela}, \citenamefont {M\"{u}ller},\ and\ \citenamefont {Tkatchenko}}]{Kabylda2025}%
  \BibitemOpen
  \bibfield  {author} {\bibinfo {author} {\bibfnamefont {Adil}\ \bibnamefont {Kabylda}}, \bibinfo {author} {\bibfnamefont {J.~Thorben}\ \bibnamefont {Frank}}, \bibinfo {author} {\bibfnamefont {ergio~Suarez}\ \bibnamefont {Dou}}, \bibinfo {author} {\bibfnamefont {Almaz}\ \bibnamefont {Khabibrakhmanov}}, \bibinfo {author} {\bibfnamefont {Leonardo~Medrano}\ \bibnamefont {Sandonas}}, \bibinfo {author} {\bibfnamefont {Oliver~T.}\ \bibnamefont {Unke}}, \bibinfo {author} {\bibfnamefont {Stefan}\ \bibnamefont {Chmiela}}, \bibinfo {author} {\bibfnamefont {Klaus-Robert}\ \bibnamefont {M\"{u}ller}}, \ and\ \bibinfo {author} {\bibfnamefont {Alexandre}\ \bibnamefont {Tkatchenko}},\ }\bibfield  {title} {\enquote {\bibinfo {title} {Molecular simulations with a pretrained neural network and universal pairwise force fields},}\ }\href {http://dx.doi.org/10.26434/chemrxiv-2024-bdfr0-v2} {\bibfield  {journal} {\bibinfo  {journal} {ChemRxiv}\ } (\bibinfo {year} {2025})}\BibitemShut {NoStop}%
\bibitem [{\citenamefont {Jacobson}\ \emph {et~al.}(2022)\citenamefont {Jacobson}, \citenamefont {Stevenson}, \citenamefont {Ramezanghorbani}, \citenamefont {Ghoreishi}, \citenamefont {Leswing}, \citenamefont {Harder},\ and\ \citenamefont {Abel}}]{Jacobson2022Transferablea}%
  \BibitemOpen
  \bibfield  {author} {\bibinfo {author} {\bibfnamefont {Leif~D.}\ \bibnamefont {Jacobson}}, \bibinfo {author} {\bibfnamefont {James~M.}\ \bibnamefont {Stevenson}}, \bibinfo {author} {\bibfnamefont {Farhad}\ \bibnamefont {Ramezanghorbani}}, \bibinfo {author} {\bibfnamefont {Delaram}\ \bibnamefont {Ghoreishi}}, \bibinfo {author} {\bibfnamefont {Karl}\ \bibnamefont {Leswing}}, \bibinfo {author} {\bibfnamefont {Edward~D.}\ \bibnamefont {Harder}}, \ and\ \bibinfo {author} {\bibfnamefont {Robert}\ \bibnamefont {Abel}},\ }\bibfield  {title} {\enquote {\bibinfo {title} {Transferable neural network potential energy surfaces for closed-shell organic molecules: Extension to ions},}\ }\href {\doibase 10.1021/acs.jctc.1c00821} {\bibfield  {journal} {\bibinfo  {journal} {Journal of Chemical Theory and Computation}\ }\textbf {\bibinfo {volume} {18}},\ \bibinfo {pages} {2354--2366} (\bibinfo {year} {2022})}\BibitemShut {NoStop}%
\bibitem [{\citenamefont {Th{\"u}rlemann}\ \emph {et~al.}(2022)\citenamefont {Th{\"u}rlemann}, \citenamefont {B{\"o}selt},\ and\ \citenamefont {Riniker}}]{Thurlemann2022Learninga}%
  \BibitemOpen
  \bibfield  {author} {\bibinfo {author} {\bibfnamefont {Moritz}\ \bibnamefont {Th{\"u}rlemann}}, \bibinfo {author} {\bibfnamefont {Lennard}\ \bibnamefont {B{\"o}selt}}, \ and\ \bibinfo {author} {\bibfnamefont {Sereina}\ \bibnamefont {Riniker}},\ }\bibfield  {title} {\enquote {\bibinfo {title} {Learning atomic multipoles: Prediction of the electrostatic potential with equivariant graph neural networks},}\ }\href {\doibase 10.1021/acs.jctc.1c01021} {\bibfield  {journal} {\bibinfo  {journal} {Journal of Chemical Theory and Computation}\ }\textbf {\bibinfo {volume} {18}},\ \bibinfo {pages} {1701--1710} (\bibinfo {year} {2022})}\BibitemShut {NoStop}%
\bibitem [{\citenamefont {Morawietz}\ \emph {et~al.}(2012)\citenamefont {Morawietz}, \citenamefont {Sharma},\ and\ \citenamefont {Behler}}]{Morawietz2012neural}%
  \BibitemOpen
  \bibfield  {author} {\bibinfo {author} {\bibfnamefont {Tobias}\ \bibnamefont {Morawietz}}, \bibinfo {author} {\bibfnamefont {Vikas}\ \bibnamefont {Sharma}}, \ and\ \bibinfo {author} {\bibfnamefont {J{\"o}rg}\ \bibnamefont {Behler}},\ }\bibfield  {title} {\enquote {\bibinfo {title} {A neural network potential-energy surface for the water dimer based on environment-dependent atomic energies and charges},}\ }\href {\doibase 10.1063/1.3682557} {\bibfield  {journal} {\bibinfo  {journal} {The Journal of Chemical Physics}\ }\textbf {\bibinfo {volume} {136}},\ \bibinfo {pages} {064103} (\bibinfo {year} {2012})}\BibitemShut {NoStop}%
\bibitem [{\citenamefont {Anstine}\ \emph {et~al.}(2025)\citenamefont {Anstine}, \citenamefont {Zubatyuk},\ and\ \citenamefont {Isayev}}]{Anstine2025}%
  \BibitemOpen
  \bibfield  {author} {\bibinfo {author} {\bibfnamefont {Dylan~M.}\ \bibnamefont {Anstine}}, \bibinfo {author} {\bibfnamefont {Roman}\ \bibnamefont {Zubatyuk}}, \ and\ \bibinfo {author} {\bibfnamefont {Olexandr}\ \bibnamefont {Isayev}},\ }\bibfield  {title} {\enquote {\bibinfo {title} {Aimnet2: a neural network potential to meet your neutral, charged, organic, and elemental-organic needs},}\ }\href {\doibase 10.1039/d4sc08572h} {\bibfield  {journal} {\bibinfo  {journal} {Chemical Science}\ }\textbf {\bibinfo {volume} {16}},\ \bibinfo {pages} {10228–10244} (\bibinfo {year} {2025})}\BibitemShut {NoStop}%
\bibitem [{\citenamefont {Rinaldi}\ \emph {et~al.}(2025)\citenamefont {Rinaldi}, \citenamefont {Bochkarev}, \citenamefont {Lysogorskiy},\ and\ \citenamefont {Drautz}}]{Rinaldi2025Chargeconstrained}%
  \BibitemOpen
  \bibfield  {author} {\bibinfo {author} {\bibfnamefont {Matteo}\ \bibnamefont {Rinaldi}}, \bibinfo {author} {\bibfnamefont {Anton}\ \bibnamefont {Bochkarev}}, \bibinfo {author} {\bibfnamefont {Yury}\ \bibnamefont {Lysogorskiy}}, \ and\ \bibinfo {author} {\bibfnamefont {Ralf}\ \bibnamefont {Drautz}},\ }\bibfield  {title} {\enquote {\bibinfo {title} {Charge-constrained atomic cluster expansion},}\ }\href {\doibase 10.1103/PhysRevMaterials.9.033802} {\bibfield  {journal} {\bibinfo  {journal} {Physical Review Materials}\ }\textbf {\bibinfo {volume} {9}},\ \bibinfo {pages} {033802} (\bibinfo {year} {2025})}\BibitemShut {NoStop}%
\bibitem [{\citenamefont {Unke}\ \emph {et~al.}(2021{\natexlab{b}})\citenamefont {Unke}, \citenamefont {Chmiela}, \citenamefont {Gastegger}, \citenamefont {Sch{\"u}tt}, \citenamefont {Sauceda},\ and\ \citenamefont {M{\"u}ller}}]{unke2021spookynet}%
  \BibitemOpen
  \bibfield  {author} {\bibinfo {author} {\bibfnamefont {Oliver~T}\ \bibnamefont {Unke}}, \bibinfo {author} {\bibfnamefont {Stefan}\ \bibnamefont {Chmiela}}, \bibinfo {author} {\bibfnamefont {Michael}\ \bibnamefont {Gastegger}}, \bibinfo {author} {\bibfnamefont {Kristof~T}\ \bibnamefont {Sch{\"u}tt}}, \bibinfo {author} {\bibfnamefont {Huziel~E}\ \bibnamefont {Sauceda}}, \ and\ \bibinfo {author} {\bibfnamefont {Klaus-Robert}\ \bibnamefont {M{\"u}ller}},\ }\bibfield  {title} {\enquote {\bibinfo {title} {Spookynet: Learning force fields with electronic degrees of freedom and nonlocal effects},}\ }\href@noop {} {\bibfield  {journal} {\bibinfo  {journal} {Nature communications}\ }\textbf {\bibinfo {volume} {12}},\ \bibinfo {pages} {7273} (\bibinfo {year} {2021}{\natexlab{b}})}\BibitemShut {NoStop}%
\bibitem [{\citenamefont {Zhang}\ \emph {et~al.}(2022)\citenamefont {Zhang}, \citenamefont {Wang}, \citenamefont {Muniz}, \citenamefont {Panagiotopoulos}, \citenamefont {Car},\ and\ \citenamefont {E}}]{zhang2022deep}%
  \BibitemOpen
  \bibfield  {author} {\bibinfo {author} {\bibfnamefont {Linfeng}\ \bibnamefont {Zhang}}, \bibinfo {author} {\bibfnamefont {Han}\ \bibnamefont {Wang}}, \bibinfo {author} {\bibfnamefont {Maria~Carolina}\ \bibnamefont {Muniz}}, \bibinfo {author} {\bibfnamefont {Athanassios~Z.}\ \bibnamefont {Panagiotopoulos}}, \bibinfo {author} {\bibfnamefont {Roberto}\ \bibnamefont {Car}}, \ and\ \bibinfo {author} {\bibfnamefont {Weinan}\ \bibnamefont {E}},\ }\bibfield  {title} {\enquote {\bibinfo {title} {A deep potential model with long-range electrostatic interactions},}\ }\href@noop {} {\bibfield  {journal} {\bibinfo  {journal} {The Journal of Chemical Physics}\ }\textbf {\bibinfo {volume} {156}} (\bibinfo {year} {2022})}\BibitemShut {NoStop}%
\bibitem [{\citenamefont {Gao}\ and\ \citenamefont {Remsing}(2022)}]{gao2022self}%
  \BibitemOpen
  \bibfield  {author} {\bibinfo {author} {\bibfnamefont {Ang}\ \bibnamefont {Gao}}\ and\ \bibinfo {author} {\bibfnamefont {Richard~C}\ \bibnamefont {Remsing}},\ }\bibfield  {title} {\enquote {\bibinfo {title} {Self-consistent determination of long-range electrostatics in neural network potentials},}\ }\href@noop {} {\bibfield  {journal} {\bibinfo  {journal} {Nature communications}\ }\textbf {\bibinfo {volume} {13}},\ \bibinfo {pages} {1572} (\bibinfo {year} {2022})}\BibitemShut {NoStop}%
\bibitem [{\citenamefont {Yu}\ \emph {et~al.}(2022)\citenamefont {Yu}, \citenamefont {Hong}, \citenamefont {Chen}, \citenamefont {Gong},\ and\ \citenamefont {Xiang}}]{yu2022capturing}%
  \BibitemOpen
  \bibfield  {author} {\bibinfo {author} {\bibfnamefont {Hongyu}\ \bibnamefont {Yu}}, \bibinfo {author} {\bibfnamefont {Liangliang}\ \bibnamefont {Hong}}, \bibinfo {author} {\bibfnamefont {Shiyou}\ \bibnamefont {Chen}}, \bibinfo {author} {\bibfnamefont {Xingao}\ \bibnamefont {Gong}}, \ and\ \bibinfo {author} {\bibfnamefont {Hongjun}\ \bibnamefont {Xiang}},\ }\bibfield  {title} {\enquote {\bibinfo {title} {Capturing long-range interaction with reciprocal space neural network},}\ }\href@noop {} {\bibfield  {journal} {\bibinfo  {journal} {arXiv preprint arXiv:2211.16684}\ } (\bibinfo {year} {2022})}\BibitemShut {NoStop}%
\bibitem [{\citenamefont {Kosmala}\ \emph {et~al.}(2023)\citenamefont {Kosmala}, \citenamefont {Gasteiger}, \citenamefont {Gao},\ and\ \citenamefont {G{\"u}nnemann}}]{kosmala2023ewald}%
  \BibitemOpen
  \bibfield  {author} {\bibinfo {author} {\bibfnamefont {Arthur}\ \bibnamefont {Kosmala}}, \bibinfo {author} {\bibfnamefont {Johannes}\ \bibnamefont {Gasteiger}}, \bibinfo {author} {\bibfnamefont {Nicholas}\ \bibnamefont {Gao}}, \ and\ \bibinfo {author} {\bibfnamefont {Stephan}\ \bibnamefont {G{\"u}nnemann}},\ }\bibfield  {title} {\enquote {\bibinfo {title} {Ewald-based long-range message passing for molecular graphs},}\ }in\ \href@noop {} {\emph {\bibinfo {booktitle} {International Conference on Machine Learning}}}\ (\bibinfo {organization} {PMLR},\ \bibinfo {year} {2023})\ pp.\ \bibinfo {pages} {17544--17563}\BibitemShut {NoStop}%
\bibitem [{\citenamefont {Grisafi}\ and\ \citenamefont {Ceriotti}(2019)}]{grisafi2019incorporating}%
  \BibitemOpen
  \bibfield  {author} {\bibinfo {author} {\bibfnamefont {Andrea}\ \bibnamefont {Grisafi}}\ and\ \bibinfo {author} {\bibfnamefont {Michele}\ \bibnamefont {Ceriotti}},\ }\bibfield  {title} {\enquote {\bibinfo {title} {Incorporating long-range physics in atomic-scale machine learning},}\ }\href@noop {} {\bibfield  {journal} {\bibinfo  {journal} {The Journal of chemical physics}\ }\textbf {\bibinfo {volume} {151}} (\bibinfo {year} {2019})}\BibitemShut {NoStop}%
\bibitem [{\citenamefont {Huguenin-Dumittan}\ \emph {et~al.}(2023)\citenamefont {Huguenin-Dumittan}, \citenamefont {Loche}, \citenamefont {Haoran},\ and\ \citenamefont {Ceriotti}}]{huguenin2023physics}%
  \BibitemOpen
  \bibfield  {author} {\bibinfo {author} {\bibfnamefont {Kevin~K}\ \bibnamefont {Huguenin-Dumittan}}, \bibinfo {author} {\bibfnamefont {Philip}\ \bibnamefont {Loche}}, \bibinfo {author} {\bibfnamefont {Ni}~\bibnamefont {Haoran}}, \ and\ \bibinfo {author} {\bibfnamefont {Michele}\ \bibnamefont {Ceriotti}},\ }\bibfield  {title} {\enquote {\bibinfo {title} {Physics-inspired equivariant descriptors of nonbonded interactions},}\ }\href@noop {} {\bibfield  {journal} {\bibinfo  {journal} {The Journal of Physical Chemistry Letters}\ }\textbf {\bibinfo {volume} {14}},\ \bibinfo {pages} {9612--9618} (\bibinfo {year} {2023})}\BibitemShut {NoStop}%
\bibitem [{\citenamefont {Faller}\ \emph {et~al.}(2024)\citenamefont {Faller}, \citenamefont {Kaltak},\ and\ \citenamefont {Kresse}}]{Faller2024Densitybased}%
  \BibitemOpen
  \bibfield  {author} {\bibinfo {author} {\bibfnamefont {Carolin}\ \bibnamefont {Faller}}, \bibinfo {author} {\bibfnamefont {Merzuk}\ \bibnamefont {Kaltak}}, \ and\ \bibinfo {author} {\bibfnamefont {Georg}\ \bibnamefont {Kresse}},\ }\bibfield  {title} {\enquote {\bibinfo {title} {Density-based long-range electrostatic descriptors for machine learning force fields},}\ }\href {\doibase 10.1063/5.0245615} {\bibfield  {journal} {\bibinfo  {journal} {The Journal of Chemical Physics}\ }\textbf {\bibinfo {volume} {161}},\ \bibinfo {pages} {214701} (\bibinfo {year} {2024})}\BibitemShut {NoStop}%
\bibitem [{\citenamefont {Monacelli}\ and\ \citenamefont {Marzari}(2024)}]{monacelli2024electrostatic}%
  \BibitemOpen
  \bibfield  {author} {\bibinfo {author} {\bibfnamefont {Lorenzo}\ \bibnamefont {Monacelli}}\ and\ \bibinfo {author} {\bibfnamefont {Nicola}\ \bibnamefont {Marzari}},\ }\bibfield  {title} {\enquote {\bibinfo {title} {Electrostatic interactions in atomistic and machine-learned potentials for polar materials},}\ }\href@noop {} {\bibfield  {journal} {\bibinfo  {journal} {arXiv preprint arXiv:2412.01642}\ } (\bibinfo {year} {2024})}\BibitemShut {NoStop}%
\bibitem [{\citenamefont {Loche}\ \emph {et~al.}(2025)\citenamefont {Loche}, \citenamefont {{Huguenin-Dumittan}}, \citenamefont {Honarmand}, \citenamefont {Xu}, \citenamefont {Rumiantsev}, \citenamefont {How}, \citenamefont {Langer},\ and\ \citenamefont {Ceriotti}}]{Loche2025Fastb}%
  \BibitemOpen
  \bibfield  {author} {\bibinfo {author} {\bibfnamefont {Philip}\ \bibnamefont {Loche}}, \bibinfo {author} {\bibfnamefont {Kevin~K.}\ \bibnamefont {{Huguenin-Dumittan}}}, \bibinfo {author} {\bibfnamefont {Melika}\ \bibnamefont {Honarmand}}, \bibinfo {author} {\bibfnamefont {Qianjun}\ \bibnamefont {Xu}}, \bibinfo {author} {\bibfnamefont {Egor}\ \bibnamefont {Rumiantsev}}, \bibinfo {author} {\bibfnamefont {Wei~Bin}\ \bibnamefont {How}}, \bibinfo {author} {\bibfnamefont {Marcel~F.}\ \bibnamefont {Langer}}, \ and\ \bibinfo {author} {\bibfnamefont {Michele}\ \bibnamefont {Ceriotti}},\ }\bibfield  {title} {\enquote {\bibinfo {title} {Fast and flexible long-range models for atomistic machine learning},}\ }\href {\doibase 10.1063/5.0251713} {\bibfield  {journal} {\bibinfo  {journal} {The Journal of Chemical Physics}\ }\textbf {\bibinfo {volume} {162}},\ \bibinfo {pages} {142501} (\bibinfo {year} {2025})}\BibitemShut {NoStop}%
\bibitem [{\citenamefont {Rumiantsev}\ \emph {et~al.}()\citenamefont {Rumiantsev}, \citenamefont {Langer}, \citenamefont {Sodjargal}, \citenamefont {Ceriotti},\ and\ \citenamefont {Loche}}]{Rumiantsev2025Learning}%
  \BibitemOpen
  \bibfield  {author} {\bibinfo {author} {\bibfnamefont {Egor}\ \bibnamefont {Rumiantsev}}, \bibinfo {author} {\bibfnamefont {Marcel~F.}\ \bibnamefont {Langer}}, \bibinfo {author} {\bibfnamefont {Tulga-Erdene}\ \bibnamefont {Sodjargal}}, \bibinfo {author} {\bibfnamefont {Michele}\ \bibnamefont {Ceriotti}}, \ and\ \bibinfo {author} {\bibfnamefont {Philip}\ \bibnamefont {Loche}},\ }\href {\doibase 10.48550/arXiv.2507.19382} {\enquote {\bibinfo {title} {Learning long-range representations with equivariant messages},}\ }\Eprint {http://arxiv.org/abs/2507.19382} {2507.19382} \BibitemShut {NoStop}%
\bibitem [{\citenamefont {Caruso}\ \emph {et~al.}(2025)\citenamefont {Caruso}, \citenamefont {Venturin}, \citenamefont {Giambagli}, \citenamefont {Rolando}, \citenamefont {No{\'e}},\ and\ \citenamefont {Clementi}}]{caruso2025Extending}%
  \BibitemOpen
  \bibfield  {author} {\bibinfo {author} {\bibfnamefont {Alessandro}\ \bibnamefont {Caruso}}, \bibinfo {author} {\bibfnamefont {Jacopo}\ \bibnamefont {Venturin}}, \bibinfo {author} {\bibfnamefont {Lorenzo}\ \bibnamefont {Giambagli}}, \bibinfo {author} {\bibfnamefont {Edoardo}\ \bibnamefont {Rolando}}, \bibinfo {author} {\bibfnamefont {Frank}\ \bibnamefont {No{\'e}}}, \ and\ \bibinfo {author} {\bibfnamefont {Cecilia}\ \bibnamefont {Clementi}},\ }\bibfield  {title} {\enquote {\bibinfo {title} {Extending the range of graph neural networks: Relaying attention nodes for global encoding},}\ }\href@noop {} {\bibfield  {journal} {\bibinfo  {journal} {arXiv preprint arXiv:2502.13797}\ } (\bibinfo {year} {2025})}\BibitemShut {NoStop}%
\bibitem [{\citenamefont {Weber}\ \emph {et~al.}(2025)\citenamefont {Weber}, \citenamefont {Guha}, \citenamefont {Agarwal}, \citenamefont {Wei}, \citenamefont {Fike}, \citenamefont {Xie}, \citenamefont {Stevenson}, \citenamefont {Leswing}, \citenamefont {Halls}, \citenamefont {Abel},\ and\ \citenamefont {Jacobson}}]{weber2025efficient}%
  \BibitemOpen
  \bibfield  {author} {\bibinfo {author} {\bibfnamefont {John~L.}\ \bibnamefont {Weber}}, \bibinfo {author} {\bibfnamefont {Rishabh~D.}\ \bibnamefont {Guha}}, \bibinfo {author} {\bibfnamefont {Garvit}\ \bibnamefont {Agarwal}}, \bibinfo {author} {\bibfnamefont {Yujing}\ \bibnamefont {Wei}}, \bibinfo {author} {\bibfnamefont {Aidan~A.}\ \bibnamefont {Fike}}, \bibinfo {author} {\bibfnamefont {Xiaowei}\ \bibnamefont {Xie}}, \bibinfo {author} {\bibfnamefont {James}\ \bibnamefont {Stevenson}}, \bibinfo {author} {\bibfnamefont {Karl}\ \bibnamefont {Leswing}}, \bibinfo {author} {\bibfnamefont {Mathew~D.}\ \bibnamefont {Halls}}, \bibinfo {author} {\bibfnamefont {Robert}\ \bibnamefont {Abel}}, \ and\ \bibinfo {author} {\bibfnamefont {Leif~D.}\ \bibnamefont {Jacobson}},\ }\bibfield  {title} {\enquote {\bibinfo {title} {Efficient long-range machine learning force fields for liquid and materials properties},}\ }\href@noop {} {\bibfield  {journal} {\bibinfo  {journal} {arXiv preprint arXiv:2505.06462}\ } (\bibinfo {year}
  {2025})}\BibitemShut {NoStop}%
\bibitem [{\citenamefont {Ghasemi}\ \emph {et~al.}(2015)\citenamefont {Ghasemi}, \citenamefont {Hofstetter}, \citenamefont {Saha},\ and\ \citenamefont {Goedecker}}]{Ghasemi2015Interatomic}%
  \BibitemOpen
  \bibfield  {author} {\bibinfo {author} {\bibfnamefont {S.~Alireza}\ \bibnamefont {Ghasemi}}, \bibinfo {author} {\bibfnamefont {Albert}\ \bibnamefont {Hofstetter}}, \bibinfo {author} {\bibfnamefont {Santanu}\ \bibnamefont {Saha}}, \ and\ \bibinfo {author} {\bibfnamefont {Stefan}\ \bibnamefont {Goedecker}},\ }\bibfield  {title} {\enquote {\bibinfo {title} {Interatomic potentials for ionic systems with density functional accuracy based on charge densities obtained by a neural network},}\ }\href {\doibase 10.1103/PhysRevB.92.045131} {\bibfield  {journal} {\bibinfo  {journal} {Physical Review B}\ }\textbf {\bibinfo {volume} {92}},\ \bibinfo {pages} {045131} (\bibinfo {year} {2015})}\BibitemShut {NoStop}%
\bibitem [{\citenamefont {Fuchs}\ \emph {et~al.}(2025)\citenamefont {Fuchs}, \citenamefont {Sanocki},\ and\ \citenamefont {Zavadlav}}]{fuchs2025Learninga}%
  \BibitemOpen
  \bibfield  {author} {\bibinfo {author} {\bibfnamefont {Paul}\ \bibnamefont {Fuchs}}, \bibinfo {author} {\bibfnamefont {Micha{\l}}\ \bibnamefont {Sanocki}}, \ and\ \bibinfo {author} {\bibfnamefont {Julija}\ \bibnamefont {Zavadlav}},\ }\bibfield  {title} {\enquote {\bibinfo {title} {Learning non-local molecular interactions via equivariant local representations and charge equilibration},}\ }\href {\doibase 10.1038/s41524-025-01790-4} {\bibfield  {journal} {\bibinfo  {journal} {npj Computational Materials}\ }\textbf {\bibinfo {volume} {11}},\ \bibinfo {pages} {287} (\bibinfo {year} {2025})}\BibitemShut {NoStop}%
\bibitem [{\citenamefont {Maruf}\ \emph {et~al.}(2025)\citenamefont {Maruf}, \citenamefont {Kim},\ and\ \citenamefont {Ahmad}}]{Maruf2025Learninga}%
  \BibitemOpen
  \bibfield  {author} {\bibinfo {author} {\bibfnamefont {Moin~Uddin}\ \bibnamefont {Maruf}}, \bibinfo {author} {\bibfnamefont {Sungmin}\ \bibnamefont {Kim}}, \ and\ \bibinfo {author} {\bibfnamefont {Zeeshan}\ \bibnamefont {Ahmad}},\ }\bibfield  {title} {\enquote {\bibinfo {title} {Learning long-range interactions in equivariant machine learning interatomic potentials via electronic degrees of freedom},}\ }\href {\doibase 10.1021/acs.jpclett.5c02352} {\bibfield  {journal} {\bibinfo  {journal} {The Journal of Physical Chemistry Letters}\ }\textbf {\bibinfo {volume} {16}},\ \bibinfo {pages} {9078--9087} (\bibinfo {year} {2025})}\BibitemShut {NoStop}%
\bibitem [{\citenamefont {Xu}\ \emph {et~al.}(2025)\citenamefont {Xu}, \citenamefont {Xie}, \citenamefont {Xie},\ and\ \citenamefont {Hu}}]{Xu2025universala}%
  \BibitemOpen
  \bibfield  {author} {\bibinfo {author} {\bibfnamefont {Zemin}\ \bibnamefont {Xu}}, \bibinfo {author} {\bibfnamefont {Wenbo}\ \bibnamefont {Xie}}, \bibinfo {author} {\bibfnamefont {Daiqian}\ \bibnamefont {Xie}}, \ and\ \bibinfo {author} {\bibfnamefont {P.}~\bibnamefont {Hu}},\ }\bibfield  {title} {\enquote {\bibinfo {title} {Towards universal property prediction in cartesian space: Tace is all you need},}\ }\href@noop {} {\bibfield  {journal} {\bibinfo  {journal} {arXiv preprint arXiv:2509.14961}\ } (\bibinfo {year} {2025})}\BibitemShut {NoStop}%
\bibitem [{\citenamefont {Pl{\'e}}\ \emph {et~al.}(2025)\citenamefont {Pl{\'e}}, \citenamefont {Adjoua}, \citenamefont {Benali}, \citenamefont {Posenitskiy}, \citenamefont {Villot}, \citenamefont {Lagard{\`e}re},\ and\ \citenamefont {Piquemal}}]{ple2025Foundation}%
  \BibitemOpen
  \bibfield  {author} {\bibinfo {author} {\bibfnamefont {Thomas}\ \bibnamefont {Pl{\'e}}}, \bibinfo {author} {\bibfnamefont {Olivier}\ \bibnamefont {Adjoua}}, \bibinfo {author} {\bibfnamefont {Anouar}\ \bibnamefont {Benali}}, \bibinfo {author} {\bibfnamefont {Evgeny}\ \bibnamefont {Posenitskiy}}, \bibinfo {author} {\bibfnamefont {Corentin}\ \bibnamefont {Villot}}, \bibinfo {author} {\bibfnamefont {Louis}\ \bibnamefont {Lagard{\`e}re}}, \ and\ \bibinfo {author} {\bibfnamefont {Jean-Philip}\ \bibnamefont {Piquemal}},\ }\bibfield  {title} {\enquote {\bibinfo {title} {A foundation model for accurate atomistic simulations in drug design},}\ }\href@noop {} {\bibfield  {journal} {\bibinfo  {journal} {ChemRxiv}\ } (\bibinfo {year} {2025})}\BibitemShut {NoStop}%
\bibitem [{\citenamefont {Gao}\ \emph {et~al.}(2025)\citenamefont {Gao}, \citenamefont {Yam}, \citenamefont {Mao}, \citenamefont {Chen}, \citenamefont {Chen},\ and\ \citenamefont {Hu}}]{Gao2025foundation}%
  \BibitemOpen
  \bibfield  {author} {\bibinfo {author} {\bibfnamefont {Rongzhi}\ \bibnamefont {Gao}}, \bibinfo {author} {\bibfnamefont {ChiYung}\ \bibnamefont {Yam}}, \bibinfo {author} {\bibfnamefont {Jianjun}\ \bibnamefont {Mao}}, \bibinfo {author} {\bibfnamefont {Shuguang}\ \bibnamefont {Chen}}, \bibinfo {author} {\bibfnamefont {GuanHua}\ \bibnamefont {Chen}}, \ and\ \bibinfo {author} {\bibfnamefont {Ziyang}\ \bibnamefont {Hu}},\ }\bibfield  {title} {\enquote {\bibinfo {title} {A foundation machine learning potential with polarizable long-range interactions for materials modelling},}\ }\href {\doibase 10.1038/s41467-025-65496-3} {\bibfield  {journal} {\bibinfo  {journal} {Nature Communications}\ }\textbf {\bibinfo {volume} {16}},\ \bibinfo {pages} {10484} (\bibinfo {year} {2025})}\BibitemShut {NoStop}%
\bibitem [{\citenamefont {Rappe}\ and\ \citenamefont {Goddard~III}(1991)}]{rappe1991charge}%
  \BibitemOpen
  \bibfield  {author} {\bibinfo {author} {\bibfnamefont {Anthony~K}\ \bibnamefont {Rappe}}\ and\ \bibinfo {author} {\bibfnamefont {William~A}\ \bibnamefont {Goddard~III}},\ }\bibfield  {title} {\enquote {\bibinfo {title} {Charge equilibration for molecular dynamics simulations},}\ }\href@noop {} {\bibfield  {journal} {\bibinfo  {journal} {The Journal of Physical Chemistry}\ }\textbf {\bibinfo {volume} {95}},\ \bibinfo {pages} {3358--3363} (\bibinfo {year} {1991})}\BibitemShut {NoStop}%
\bibitem [{\citenamefont {Falletta}\ \emph {et~al.}(2025)\citenamefont {Falletta}, \citenamefont {Cepellotti}, \citenamefont {Johansson}, \citenamefont {Tan}, \citenamefont {Descoteaux}, \citenamefont {Musaelian}, \citenamefont {Owen},\ and\ \citenamefont {Kozinsky}}]{falletta2025unified}%
  \BibitemOpen
  \bibfield  {author} {\bibinfo {author} {\bibfnamefont {Stefano}\ \bibnamefont {Falletta}}, \bibinfo {author} {\bibfnamefont {Andrea}\ \bibnamefont {Cepellotti}}, \bibinfo {author} {\bibfnamefont {Anders}\ \bibnamefont {Johansson}}, \bibinfo {author} {\bibfnamefont {Chuin~Wei}\ \bibnamefont {Tan}}, \bibinfo {author} {\bibfnamefont {Marc~L}\ \bibnamefont {Descoteaux}}, \bibinfo {author} {\bibfnamefont {Albert}\ \bibnamefont {Musaelian}}, \bibinfo {author} {\bibfnamefont {Cameron~J}\ \bibnamefont {Owen}}, \ and\ \bibinfo {author} {\bibfnamefont {Boris}\ \bibnamefont {Kozinsky}},\ }\bibfield  {title} {\enquote {\bibinfo {title} {Unified differentiable learning of electric response},}\ }\href@noop {} {\bibfield  {journal} {\bibinfo  {journal} {Nature communications}\ }\textbf {\bibinfo {volume} {16}},\ \bibinfo {pages} {4031} (\bibinfo {year} {2025})}\BibitemShut {NoStop}%
\bibitem [{\citenamefont {Schmiedmayer}\ and\ \citenamefont {Kresse}(2024)}]{Schmiedmayer2024}%
  \BibitemOpen
  \bibfield  {author} {\bibinfo {author} {\bibfnamefont {Bernhard}\ \bibnamefont {Schmiedmayer}}\ and\ \bibinfo {author} {\bibfnamefont {Georg}\ \bibnamefont {Kresse}},\ }\bibfield  {title} {\enquote {\bibinfo {title} {Derivative learning of tensorial quantities—predicting finite temperature infrared spectra from first principles},}\ }\href {http://dx.doi.org/10.1063/5.0217243} {\bibfield  {journal} {\bibinfo  {journal} {The Journal of Chemical Physics}\ }\textbf {\bibinfo {volume} {161}} (\bibinfo {year} {2024})}\BibitemShut {NoStop}%
\bibitem [{\citenamefont {Stocco}\ \emph {et~al.}(2025)\citenamefont {Stocco}, \citenamefont {Carbogno},\ and\ \citenamefont {Rossi}}]{Stocco2025Electricfield}%
  \BibitemOpen
  \bibfield  {author} {\bibinfo {author} {\bibfnamefont {Elia}\ \bibnamefont {Stocco}}, \bibinfo {author} {\bibfnamefont {Christian}\ \bibnamefont {Carbogno}}, \ and\ \bibinfo {author} {\bibfnamefont {Mariana}\ \bibnamefont {Rossi}},\ }\bibfield  {title} {\enquote {\bibinfo {title} {Electric-field driven nuclear dynamics of liquids and solids from a multi-valued machine-learned dipolar model},}\ }\href {\doibase 10.1038/s41524-025-01751-x} {\bibfield  {journal} {\bibinfo  {journal} {npj Computational Materials}\ }\textbf {\bibinfo {volume} {11}},\ \bibinfo {pages} {304} (\bibinfo {year} {2025})}\BibitemShut {NoStop}%
\bibitem [{\citenamefont {Zhang}\ and\ \citenamefont {Jiang}(2023)}]{Zhang2023}%
  \BibitemOpen
  \bibfield  {author} {\bibinfo {author} {\bibfnamefont {Yaolong}\ \bibnamefont {Zhang}}\ and\ \bibinfo {author} {\bibfnamefont {Bin}\ \bibnamefont {Jiang}},\ }\bibfield  {title} {\enquote {\bibinfo {title} {Universal machine learning for the response of atomistic systems to external fields},}\ }\href {\doibase 10.1038/s41467-023-42148-y} {\bibfield  {journal} {\bibinfo  {journal} {Nature Communications}\ }\textbf {\bibinfo {volume} {14}} (\bibinfo {year} {2023}),\ 10.1038/s41467-023-42148-y}\BibitemShut {NoStop}%
\bibitem [{\citenamefont {Gastegger}\ \emph {et~al.}(2021)\citenamefont {Gastegger}, \citenamefont {Sch\"{u}tt},\ and\ \citenamefont {M\"{u}ller}}]{Gastegger2021}%
  \BibitemOpen
  \bibfield  {author} {\bibinfo {author} {\bibfnamefont {Michael}\ \bibnamefont {Gastegger}}, \bibinfo {author} {\bibfnamefont {Kristof~T.}\ \bibnamefont {Sch\"{u}tt}}, \ and\ \bibinfo {author} {\bibfnamefont {Klaus-Robert}\ \bibnamefont {M\"{u}ller}},\ }\bibfield  {title} {\enquote {\bibinfo {title} {Machine learning of solvent effects on molecular spectra and reactions},}\ }\href {\doibase 10.1039/d1sc02742e} {\bibfield  {journal} {\bibinfo  {journal} {Chemical Science}\ }\textbf {\bibinfo {volume} {12}},\ \bibinfo {pages} {11473–11483} (\bibinfo {year} {2021})}\BibitemShut {NoStop}%
\bibitem [{\citenamefont {King}\ \emph {et~al.}(2025)\citenamefont {King}, \citenamefont {Kim}, \citenamefont {Zhong},\ and\ \citenamefont {Cheng}}]{King2025Machine}%
  \BibitemOpen
  \bibfield  {author} {\bibinfo {author} {\bibfnamefont {Daniel~S.}\ \bibnamefont {King}}, \bibinfo {author} {\bibfnamefont {Dongjin}\ \bibnamefont {Kim}}, \bibinfo {author} {\bibfnamefont {Peichen}\ \bibnamefont {Zhong}}, \ and\ \bibinfo {author} {\bibfnamefont {Bingqing}\ \bibnamefont {Cheng}},\ }\bibfield  {title} {\enquote {\bibinfo {title} {Machine learning of charges and long-range interactions from energies and forces},}\ }\href {\doibase 10.1038/s41467-025-63852-x} {\bibfield  {journal} {\bibinfo  {journal} {Nature Communications}\ }\textbf {\bibinfo {volume} {16}},\ \bibinfo {pages} {8763} (\bibinfo {year} {2025})}\BibitemShut {NoStop}%
\bibitem [{\citenamefont {Zhong}\ \emph {et~al.}(2025)\citenamefont {Zhong}, \citenamefont {Kim}, \citenamefont {King},\ and\ \citenamefont {Cheng}}]{zhong2025machine}%
  \BibitemOpen
  \bibfield  {author} {\bibinfo {author} {\bibfnamefont {Peichen}\ \bibnamefont {Zhong}}, \bibinfo {author} {\bibfnamefont {Dongjin}\ \bibnamefont {Kim}}, \bibinfo {author} {\bibfnamefont {Daniel~S}\ \bibnamefont {King}}, \ and\ \bibinfo {author} {\bibfnamefont {Bingqing}\ \bibnamefont {Cheng}},\ }\bibfield  {title} {\enquote {\bibinfo {title} {Machine learning interatomic potential can infer electrical response},}\ }\href@noop {} {\bibfield  {journal} {\bibinfo  {journal} {arXiv preprint arXiv:2504.05169}\ } (\bibinfo {year} {2025})}\BibitemShut {NoStop}%
\bibitem [{\citenamefont {Kim}\ \emph {et~al.}(2025)\citenamefont {Kim}, \citenamefont {Wang}, \citenamefont {Vargas}, \citenamefont {Zhong}, \citenamefont {King}, \citenamefont {Inizan},\ and\ \citenamefont {Cheng}}]{Kim2025Universalb}%
  \BibitemOpen
  \bibfield  {author} {\bibinfo {author} {\bibfnamefont {Dongjin}\ \bibnamefont {Kim}}, \bibinfo {author} {\bibfnamefont {Xiaoyu}\ \bibnamefont {Wang}}, \bibinfo {author} {\bibfnamefont {Santiago}\ \bibnamefont {Vargas}}, \bibinfo {author} {\bibfnamefont {Peichen}\ \bibnamefont {Zhong}}, \bibinfo {author} {\bibfnamefont {Daniel~S.}\ \bibnamefont {King}}, \bibinfo {author} {\bibfnamefont {Theo~Jaffrelot}\ \bibnamefont {Inizan}}, \ and\ \bibinfo {author} {\bibfnamefont {Bingqing}\ \bibnamefont {Cheng}},\ }\bibfield  {title} {\enquote {\bibinfo {title} {A universal augmentation framework for long-range electrostatics in machine learning interatomic potentials},}\ }\href {\doibase 10.1021/acs.jctc.5c01400} {\bibfield  {journal} {\bibinfo  {journal} {Journal of Chemical Theory and Computation}\ } (\bibinfo {year} {2025}),\ 10.1021/acs.jctc.5c01400}\BibitemShut {NoStop}%
\bibitem [{\citenamefont {Grisafi}\ \emph {et~al.}(2019)\citenamefont {Grisafi}, \citenamefont {Fabrizio}, \citenamefont {Meyer}, \citenamefont {Wilkins}, \citenamefont {Corminboeuf},\ and\ \citenamefont {Ceriotti}}]{Grisafi2019Transferable}%
  \BibitemOpen
  \bibfield  {author} {\bibinfo {author} {\bibfnamefont {Andrea}\ \bibnamefont {Grisafi}}, \bibinfo {author} {\bibfnamefont {Alberto}\ \bibnamefont {Fabrizio}}, \bibinfo {author} {\bibfnamefont {Benjamin}\ \bibnamefont {Meyer}}, \bibinfo {author} {\bibfnamefont {David~M.}\ \bibnamefont {Wilkins}}, \bibinfo {author} {\bibfnamefont {Clemence}\ \bibnamefont {Corminboeuf}}, \ and\ \bibinfo {author} {\bibfnamefont {Michele}\ \bibnamefont {Ceriotti}},\ }\bibfield  {title} {\enquote {\bibinfo {title} {Transferable machine-learning model of the electron density},}\ }\href {\doibase 10.1021/acscentsci.8b00551} {\bibfield  {journal} {\bibinfo  {journal} {ACS Central Science}\ }\textbf {\bibinfo {volume} {5}},\ \bibinfo {pages} {57--64} (\bibinfo {year} {2019})}\BibitemShut {NoStop}%
\bibitem [{\citenamefont {Lewis}\ \emph {et~al.}(2021)\citenamefont {Lewis}, \citenamefont {Grisafi}, \citenamefont {Ceriotti},\ and\ \citenamefont {Rossi}}]{Lewis2021Learning}%
  \BibitemOpen
  \bibfield  {author} {\bibinfo {author} {\bibfnamefont {Alan~M.}\ \bibnamefont {Lewis}}, \bibinfo {author} {\bibfnamefont {Andrea}\ \bibnamefont {Grisafi}}, \bibinfo {author} {\bibfnamefont {Michele}\ \bibnamefont {Ceriotti}}, \ and\ \bibinfo {author} {\bibfnamefont {Mariana}\ \bibnamefont {Rossi}},\ }\bibfield  {title} {\enquote {\bibinfo {title} {Learning electron densities in the condensed phase},}\ }\href {\doibase 10.1021/acs.jctc.1c00576} {\bibfield  {journal} {\bibinfo  {journal} {Journal of Chemical Theory and Computation}\ }\textbf {\bibinfo {volume} {17}},\ \bibinfo {pages} {7203--7214} (\bibinfo {year} {2021})}\BibitemShut {NoStop}%
\bibitem [{\citenamefont {Grisafi}\ \emph {et~al.}(2023)\citenamefont {Grisafi}, \citenamefont {Bussy}, \citenamefont {Salanne},\ and\ \citenamefont {Vuilleumier}}]{grisafi2023predicting}%
  \BibitemOpen
  \bibfield  {author} {\bibinfo {author} {\bibfnamefont {Andrea}\ \bibnamefont {Grisafi}}, \bibinfo {author} {\bibfnamefont {Augustin}\ \bibnamefont {Bussy}}, \bibinfo {author} {\bibfnamefont {Mathieu}\ \bibnamefont {Salanne}}, \ and\ \bibinfo {author} {\bibfnamefont {Rodolphe}\ \bibnamefont {Vuilleumier}},\ }\bibfield  {title} {\enquote {\bibinfo {title} {Predicting the charge density response in metal electrodes},}\ }\href@noop {} {\bibfield  {journal} {\bibinfo  {journal} {Physical Review Materials}\ }\textbf {\bibinfo {volume} {7}},\ \bibinfo {pages} {125403} (\bibinfo {year} {2023})}\BibitemShut {NoStop}%
\bibitem [{\citenamefont {Yu}\ \emph {et~al.}(2025)\citenamefont {Yu}, \citenamefont {Deng}, \citenamefont {Zhu}, \citenamefont {Xie}, \citenamefont {Zhang}, \citenamefont {Shi}, \citenamefont {Zhong}, \citenamefont {He},\ and\ \citenamefont {Xiang}}]{yu_prediction_2025}%
  \BibitemOpen
  \bibfield  {author} {\bibinfo {author} {\bibfnamefont {Hongyu}\ \bibnamefont {Yu}}, \bibinfo {author} {\bibfnamefont {Shihan}\ \bibnamefont {Deng}}, \bibinfo {author} {\bibfnamefont {Haiyan}\ \bibnamefont {Zhu}}, \bibinfo {author} {\bibfnamefont {Muting}\ \bibnamefont {Xie}}, \bibinfo {author} {\bibfnamefont {Yuwen}\ \bibnamefont {Zhang}}, \bibinfo {author} {\bibfnamefont {Xizhi}\ \bibnamefont {Shi}}, \bibinfo {author} {\bibfnamefont {Jianxin}\ \bibnamefont {Zhong}}, \bibinfo {author} {\bibfnamefont {Chaoyu}\ \bibnamefont {He}}, \ and\ \bibinfo {author} {\bibfnamefont {Hongjun}\ \bibnamefont {Xiang}},\ }\bibfield  {title} {\enquote {\bibinfo {title} {Prediction of {Room} {Temperature} {Ferroelectricity} in {Subnano} {Silicon} {Thin} {Films} with an {Antiferroelectric} {Ground} {State}},}\ }\href {\doibase 10.1103/xlkt-4wk5} {\bibfield  {journal} {\bibinfo  {journal} {Physical Review Letters}\ }\textbf {\bibinfo {volume} {135}},\ \bibinfo {pages} {156801} (\bibinfo {year} {2025})}\BibitemShut {NoStop}%
\bibitem [{\citenamefont {Grisafi}\ \emph {et~al.}(2018)\citenamefont {Grisafi}, \citenamefont {Wilkins}, \citenamefont {Cs{\'a}nyi},\ and\ \citenamefont {Ceriotti}}]{grisafi2018symmetry}%
  \BibitemOpen
  \bibfield  {author} {\bibinfo {author} {\bibfnamefont {Andrea}\ \bibnamefont {Grisafi}}, \bibinfo {author} {\bibfnamefont {David~M}\ \bibnamefont {Wilkins}}, \bibinfo {author} {\bibfnamefont {G{\'a}bor}\ \bibnamefont {Cs{\'a}nyi}}, \ and\ \bibinfo {author} {\bibfnamefont {Michele}\ \bibnamefont {Ceriotti}},\ }\bibfield  {title} {\enquote {\bibinfo {title} {Symmetry-adapted machine learning for tensorial properties of atomistic systems},}\ }\href@noop {} {\bibfield  {journal} {\bibinfo  {journal} {Physical review letters}\ }\textbf {\bibinfo {volume} {120}},\ \bibinfo {pages} {036002} (\bibinfo {year} {2018})}\BibitemShut {NoStop}%
\bibitem [{\citenamefont {Schienbein}(2023)}]{schienbein2023Spectroscopy}%
  \BibitemOpen
  \bibfield  {author} {\bibinfo {author} {\bibfnamefont {Philipp}\ \bibnamefont {Schienbein}},\ }\bibfield  {title} {\enquote {\bibinfo {title} {Spectroscopy from machine learning by accurately representing the atomic polar tensor},}\ }\href {\doibase 10.1021/acs.jctc.2c00788} {\bibfield  {journal} {\bibinfo  {journal} {Journal of Chemical Theory and Computation}\ }\textbf {\bibinfo {volume} {19}},\ \bibinfo {pages} {705--712} (\bibinfo {year} {2023})}\BibitemShut {NoStop}%
\bibitem [{\citenamefont {Joll}\ \emph {et~al.}(2024)\citenamefont {Joll}, \citenamefont {Schienbein}, \citenamefont {Rosso},\ and\ \citenamefont {Blumberger}}]{Joll2024}%
  \BibitemOpen
  \bibfield  {author} {\bibinfo {author} {\bibfnamefont {Kit}\ \bibnamefont {Joll}}, \bibinfo {author} {\bibfnamefont {Philipp}\ \bibnamefont {Schienbein}}, \bibinfo {author} {\bibfnamefont {Kevin~M.}\ \bibnamefont {Rosso}}, \ and\ \bibinfo {author} {\bibfnamefont {Jochen}\ \bibnamefont {Blumberger}},\ }\bibfield  {title} {\enquote {\bibinfo {title} {Machine learning the electric field response of condensed phase systems using perturbed neural network potentials},}\ }\href {\doibase 10.1038/s41467-024-52491-3} {\bibfield  {journal} {\bibinfo  {journal} {Nature Communications}\ }\textbf {\bibinfo {volume} {15}} (\bibinfo {year} {2024}),\ 10.1038/s41467-024-52491-3}\BibitemShut {NoStop}%
\bibitem [{\citenamefont {Musaelian}\ \emph {et~al.}(2023)\citenamefont {Musaelian}, \citenamefont {Batzner}, \citenamefont {Johansson}, \citenamefont {Sun}, \citenamefont {Owen}, \citenamefont {Kornbluth},\ and\ \citenamefont {Kozinsky}}]{musaelian2023learning}%
  \BibitemOpen
  \bibfield  {author} {\bibinfo {author} {\bibfnamefont {Albert}\ \bibnamefont {Musaelian}}, \bibinfo {author} {\bibfnamefont {Simon}\ \bibnamefont {Batzner}}, \bibinfo {author} {\bibfnamefont {Anders}\ \bibnamefont {Johansson}}, \bibinfo {author} {\bibfnamefont {Lixin}\ \bibnamefont {Sun}}, \bibinfo {author} {\bibfnamefont {Cameron~J}\ \bibnamefont {Owen}}, \bibinfo {author} {\bibfnamefont {Mordechai}\ \bibnamefont {Kornbluth}}, \ and\ \bibinfo {author} {\bibfnamefont {Boris}\ \bibnamefont {Kozinsky}},\ }\bibfield  {title} {\enquote {\bibinfo {title} {Learning local equivariant representations for large-scale atomistic dynamics},}\ }\href@noop {} {\bibfield  {journal} {\bibinfo  {journal} {Nature Communications}\ }\textbf {\bibinfo {volume} {14}},\ \bibinfo {pages} {579} (\bibinfo {year} {2023})}\BibitemShut {NoStop}%
\bibitem [{\citenamefont {Cheng}(2024)}]{cheng2024cartesian}%
  \BibitemOpen
  \bibfield  {author} {\bibinfo {author} {\bibfnamefont {Bingqing}\ \bibnamefont {Cheng}},\ }\bibfield  {title} {\enquote {\bibinfo {title} {Cartesian atomic cluster expansion for machine learning interatomic potentials},}\ }\href@noop {} {\bibfield  {journal} {\bibinfo  {journal} {npj Computational Materials}\ }\textbf {\bibinfo {volume} {10}},\ \bibinfo {pages} {157} (\bibinfo {year} {2024})}\BibitemShut {NoStop}%
\bibitem [{\citenamefont {Deng}\ \emph {et~al.}(2023)\citenamefont {Deng}, \citenamefont {Zhong}, \citenamefont {Jun}, \citenamefont {Riebesell}, \citenamefont {Han}, \citenamefont {Bartel},\ and\ \citenamefont {Ceder}}]{deng2023chgnet}%
  \BibitemOpen
  \bibfield  {author} {\bibinfo {author} {\bibfnamefont {Bowen}\ \bibnamefont {Deng}}, \bibinfo {author} {\bibfnamefont {Peichen}\ \bibnamefont {Zhong}}, \bibinfo {author} {\bibfnamefont {KyuJung}\ \bibnamefont {Jun}}, \bibinfo {author} {\bibfnamefont {Janosh}\ \bibnamefont {Riebesell}}, \bibinfo {author} {\bibfnamefont {Kevin}\ \bibnamefont {Han}}, \bibinfo {author} {\bibfnamefont {Christopher~J}\ \bibnamefont {Bartel}}, \ and\ \bibinfo {author} {\bibfnamefont {Gerbrand}\ \bibnamefont {Ceder}},\ }\bibfield  {title} {\enquote {\bibinfo {title} {Chgnet as a pretrained universal neural network potential for charge-informed atomistic modelling},}\ }\href@noop {} {\bibfield  {journal} {\bibinfo  {journal} {Nature Machine Intelligence}\ }\textbf {\bibinfo {volume} {5}},\ \bibinfo {pages} {1031--1041} (\bibinfo {year} {2023})}\BibitemShut {NoStop}%
\bibitem [{\citenamefont {Wang}\ \emph {et~al.}(2025)\citenamefont {Wang}, \citenamefont {Chen}, \citenamefont {Zeng}, \citenamefont {Stein}, \citenamefont {Lim},\ and\ \citenamefont {Cheng}}]{Wang2025Ionmodulateda}%
  \BibitemOpen
  \bibfield  {author} {\bibinfo {author} {\bibfnamefont {Xiaoyu}\ \bibnamefont {Wang}}, \bibinfo {author} {\bibfnamefont {Junmin}\ \bibnamefont {Chen}}, \bibinfo {author} {\bibfnamefont {Zezhu}\ \bibnamefont {Zeng}}, \bibinfo {author} {\bibfnamefont {Frederick}\ \bibnamefont {Stein}}, \bibinfo {author} {\bibfnamefont {Junho}\ \bibnamefont {Lim}}, \ and\ \bibinfo {author} {\bibfnamefont {Bingqing}\ \bibnamefont {Cheng}},\ }\bibfield  {title} {\enquote {\bibinfo {title} {Ion-modulated structure, proton transfer, and capacitance in the pt(111)/water electric double layer},}\ }\href@noop {} {\bibfield  {journal} {\bibinfo  {journal} {arXiv preprint arXiv:2509.13727}\ } (\bibinfo {year} {2025})}\BibitemShut {NoStop}%
\bibitem [{\citenamefont {Hjorth~Larsen}\ \emph {et~al.}(2017)\citenamefont {Hjorth~Larsen}, \citenamefont {J{\o}rgen~Mortensen}, \citenamefont {Blomqvist}, \citenamefont {Castelli}, \citenamefont {Christensen}, \citenamefont {Du{\l}ak}, \citenamefont {Friis}, \citenamefont {Groves}, \citenamefont {Hammer}, \citenamefont {Hargus}, \citenamefont {Hermes}, \citenamefont {Jennings}, \citenamefont {Bjerre~Jensen}, \citenamefont {Kermode}, \citenamefont {Kitchin}, \citenamefont {Leonhard~Kolsbjerg}, \citenamefont {Kubal}, \citenamefont {Kaasbjerg}, \citenamefont {Lysgaard}, \citenamefont {Bergmann~Maronsson}, \citenamefont {Maxson}, \citenamefont {Olsen}, \citenamefont {Pastewka}, \citenamefont {Peterson}, \citenamefont {Rostgaard}, \citenamefont {Schi{\o}tz}, \citenamefont {Sch{\"u}tt}, \citenamefont {Strange}, \citenamefont {Thygesen}, \citenamefont {Vegge}, \citenamefont {Vilhelmsen}, \citenamefont {Walter}, \citenamefont {Zeng},\ and\ \citenamefont {Jacobsen}}]{HjorthLarsen2017atomic}%
  \BibitemOpen
  \bibfield  {author} {\bibinfo {author} {\bibfnamefont {Ask}\ \bibnamefont {Hjorth~Larsen}}, \bibinfo {author} {\bibfnamefont {Jens}\ \bibnamefont {J{\o}rgen~Mortensen}}, \bibinfo {author} {\bibfnamefont {Jakob}\ \bibnamefont {Blomqvist}}, \bibinfo {author} {\bibfnamefont {Ivano~E}\ \bibnamefont {Castelli}}, \bibinfo {author} {\bibfnamefont {Rune}\ \bibnamefont {Christensen}}, \bibinfo {author} {\bibfnamefont {Marcin}\ \bibnamefont {Du{\l}ak}}, \bibinfo {author} {\bibfnamefont {Jesper}\ \bibnamefont {Friis}}, \bibinfo {author} {\bibfnamefont {Michael~N}\ \bibnamefont {Groves}}, \bibinfo {author} {\bibfnamefont {Bj{\o}rk}\ \bibnamefont {Hammer}}, \bibinfo {author} {\bibfnamefont {Cory}\ \bibnamefont {Hargus}}, \bibinfo {author} {\bibfnamefont {Eric~D}\ \bibnamefont {Hermes}}, \bibinfo {author} {\bibfnamefont {Paul~C}\ \bibnamefont {Jennings}}, \bibinfo {author} {\bibfnamefont {Peter}\ \bibnamefont {Bjerre~Jensen}}, \bibinfo {author} {\bibfnamefont {James}\ \bibnamefont {Kermode}}, \bibinfo {author}
  {\bibfnamefont {John~R}\ \bibnamefont {Kitchin}}, \bibinfo {author} {\bibfnamefont {Esben}\ \bibnamefont {Leonhard~Kolsbjerg}}, \bibinfo {author} {\bibfnamefont {Joseph}\ \bibnamefont {Kubal}}, \bibinfo {author} {\bibfnamefont {Kristen}\ \bibnamefont {Kaasbjerg}}, \bibinfo {author} {\bibfnamefont {Steen}\ \bibnamefont {Lysgaard}}, \bibinfo {author} {\bibfnamefont {J{\'o}n}\ \bibnamefont {Bergmann~Maronsson}}, \bibinfo {author} {\bibfnamefont {Tristan}\ \bibnamefont {Maxson}}, \bibinfo {author} {\bibfnamefont {Thomas}\ \bibnamefont {Olsen}}, \bibinfo {author} {\bibfnamefont {Lars}\ \bibnamefont {Pastewka}}, \bibinfo {author} {\bibfnamefont {Andrew}\ \bibnamefont {Peterson}}, \bibinfo {author} {\bibfnamefont {Carsten}\ \bibnamefont {Rostgaard}}, \bibinfo {author} {\bibfnamefont {Jakob}\ \bibnamefont {Schi{\o}tz}}, \bibinfo {author} {\bibfnamefont {Ole}\ \bibnamefont {Sch{\"u}tt}}, \bibinfo {author} {\bibfnamefont {Mikkel}\ \bibnamefont {Strange}}, \bibinfo {author} {\bibfnamefont {Kristian~S}\ \bibnamefont
  {Thygesen}}, \bibinfo {author} {\bibfnamefont {Tejs}\ \bibnamefont {Vegge}}, \bibinfo {author} {\bibfnamefont {Lasse}\ \bibnamefont {Vilhelmsen}}, \bibinfo {author} {\bibfnamefont {Michael}\ \bibnamefont {Walter}}, \bibinfo {author} {\bibfnamefont {Zhenhua}\ \bibnamefont {Zeng}}, \ and\ \bibinfo {author} {\bibfnamefont {Karsten~W}\ \bibnamefont {Jacobsen}},\ }\bibfield  {title} {\enquote {\bibinfo {title} {The atomic simulation environment---a python library for working with atoms},}\ }\href {\doibase 10.1088/1361-648X/aa680e} {\bibfield  {journal} {\bibinfo  {journal} {Journal of Physics: Condensed Matter}\ }\textbf {\bibinfo {volume} {29}},\ \bibinfo {pages} {273002} (\bibinfo {year} {2017})}\BibitemShut {NoStop}%
\bibitem [{\citenamefont {Bertie}\ and\ \citenamefont {Lan}(1996)}]{Bertie1996Infrared}%
  \BibitemOpen
  \bibfield  {author} {\bibinfo {author} {\bibfnamefont {John~E.}\ \bibnamefont {Bertie}}\ and\ \bibinfo {author} {\bibfnamefont {Zhida}\ \bibnamefont {Lan}},\ }\bibfield  {title} {\enquote {\bibinfo {title} {Infrared intensities of liquids xx: The intensity of the oh stretching band of liquid water revisited, and the best current values of the optical constants of h2o(l) at 25{$^\circ$}c between 15,000 and 1 cm-1},}\ }\href {\doibase 10.1366/0003702963905385} {\bibfield  {journal} {\bibinfo  {journal} {Applied Spectroscopy}\ }\textbf {\bibinfo {volume} {50}},\ \bibinfo {pages} {1047--1057} (\bibinfo {year} {1996})}\BibitemShut {NoStop}%
\bibitem [{\citenamefont {Sch{\"u}tt}\ \emph {et~al.}(2021)\citenamefont {Sch{\"u}tt}, \citenamefont {Unke},\ and\ \citenamefont {Gastegger}}]{painn}%
  \BibitemOpen
  \bibfield  {author} {\bibinfo {author} {\bibfnamefont {Kristof}\ \bibnamefont {Sch{\"u}tt}}, \bibinfo {author} {\bibfnamefont {Oliver}\ \bibnamefont {Unke}}, \ and\ \bibinfo {author} {\bibfnamefont {Michael}\ \bibnamefont {Gastegger}},\ }\bibfield  {title} {\enquote {\bibinfo {title} {Equivariant message passing for the prediction of tensorial properties and molecular spectra},}\ }in\ \href {https://proceedings.mlr.press/v139/schutt21a.html} {\emph {\bibinfo {booktitle} {Proceedings of the 38th International Conference on Machine Learning}}},\ \bibinfo {series} {Proceedings of Machine Learning Research}, Vol.\ \bibinfo {volume} {139},\ \bibinfo {editor} {edited by\ \bibinfo {editor} {\bibfnamefont {Marina}\ \bibnamefont {Meila}}\ and\ \bibinfo {editor} {\bibfnamefont {Tong}\ \bibnamefont {Zhang}}}\ (\bibinfo  {publisher} {PMLR},\ \bibinfo {year} {2021})\ pp.\ \bibinfo {pages} {9377--9388}\BibitemShut {NoStop}%
\bibitem [{\citenamefont {Stuke}\ \emph {et~al.}(2020)\citenamefont {Stuke}, \citenamefont {Kunkel}, \citenamefont {Golze}, \citenamefont {Todorovi{\'c}}, \citenamefont {Margraf}, \citenamefont {Reuter}, \citenamefont {Rinke},\ and\ \citenamefont {Oberhofer}}]{Stuke2020Atomic}%
  \BibitemOpen
  \bibfield  {author} {\bibinfo {author} {\bibfnamefont {Annika}\ \bibnamefont {Stuke}}, \bibinfo {author} {\bibfnamefont {Christian}\ \bibnamefont {Kunkel}}, \bibinfo {author} {\bibfnamefont {Dorothea}\ \bibnamefont {Golze}}, \bibinfo {author} {\bibfnamefont {Milica}\ \bibnamefont {Todorovi{\'c}}}, \bibinfo {author} {\bibfnamefont {Johannes~T.}\ \bibnamefont {Margraf}}, \bibinfo {author} {\bibfnamefont {Karsten}\ \bibnamefont {Reuter}}, \bibinfo {author} {\bibfnamefont {Patrick}\ \bibnamefont {Rinke}}, \ and\ \bibinfo {author} {\bibfnamefont {Harald}\ \bibnamefont {Oberhofer}},\ }\bibfield  {title} {\enquote {\bibinfo {title} {Atomic structures and orbital energies of 61,489 crystal-forming organic molecules},}\ }\href {\doibase 10.1038/s41597-020-0385-y} {\bibfield  {journal} {\bibinfo  {journal} {Scientific Data}\ }\textbf {\bibinfo {volume} {7}},\ \bibinfo {pages} {58} (\bibinfo {year} {2020})}\BibitemShut {NoStop}%
\bibitem [{\citenamefont {Ko}\ \emph {et~al.}(2025)\citenamefont {Ko}, \citenamefont {Liu}, \citenamefont {Mishra}, \citenamefont {Yu}, \citenamefont {Qi},\ and\ \citenamefont {Ong}}]{Ko2025Fast}%
  \BibitemOpen
  \bibfield  {author} {\bibinfo {author} {\bibfnamefont {Tsz~Wai}\ \bibnamefont {Ko}}, \bibinfo {author} {\bibfnamefont {Runze}\ \bibnamefont {Liu}}, \bibinfo {author} {\bibfnamefont {Adesh~Rohan}\ \bibnamefont {Mishra}}, \bibinfo {author} {\bibfnamefont {Zihan}\ \bibnamefont {Yu}}, \bibinfo {author} {\bibfnamefont {Ji}~\bibnamefont {Qi}}, \ and\ \bibinfo {author} {\bibfnamefont {Shyue~Ping}\ \bibnamefont {Ong}},\ }\bibfield  {title} {\enquote {\bibinfo {title} {A fast, accurate, and reactive equivariant foundation potential},}\ }\href@noop {} {\  (\bibinfo {year} {2025})}\BibitemShut {NoStop}%
\bibitem [{\citenamefont {Semelak}\ \emph {et~al.}(2025)\citenamefont {Semelak}, \citenamefont {Pickering}, \citenamefont {Huddleston}, \citenamefont {Olmos}, \citenamefont {Grassano}, \citenamefont {Clemente}, \citenamefont {Drusin}, \citenamefont {Marti}, \citenamefont {Gonzalez~Lebrero}, \citenamefont {Roitberg},\ and\ \citenamefont {Estrin}}]{Semelak2025Advancing}%
  \BibitemOpen
  \bibfield  {author} {\bibinfo {author} {\bibfnamefont {Jonathan~A.}\ \bibnamefont {Semelak}}, \bibinfo {author} {\bibfnamefont {Ignacio}\ \bibnamefont {Pickering}}, \bibinfo {author} {\bibfnamefont {Kate}\ \bibnamefont {Huddleston}}, \bibinfo {author} {\bibfnamefont {Justo}\ \bibnamefont {Olmos}}, \bibinfo {author} {\bibfnamefont {Juan~Santiago}\ \bibnamefont {Grassano}}, \bibinfo {author} {\bibfnamefont {Camila~Mara}\ \bibnamefont {Clemente}}, \bibinfo {author} {\bibfnamefont {Salvador~I.}\ \bibnamefont {Drusin}}, \bibinfo {author} {\bibfnamefont {Marcelo}\ \bibnamefont {Marti}}, \bibinfo {author} {\bibfnamefont {Mariano~Camilo}\ \bibnamefont {Gonzalez~Lebrero}}, \bibinfo {author} {\bibfnamefont {Adrian~E.}\ \bibnamefont {Roitberg}}, \ and\ \bibinfo {author} {\bibfnamefont {Dario~A.}\ \bibnamefont {Estrin}},\ }\bibfield  {title} {\enquote {\bibinfo {title} {Advancing multiscale molecular modeling with machine learning-derived electrostatics},}\ }\href {\doibase 10.1021/acs.jctc.4c01792} {\bibfield
  {journal} {\bibinfo  {journal} {Journal of Chemical Theory and Computation}\ }\textbf {\bibinfo {volume} {21}},\ \bibinfo {pages} {5194--5207} (\bibinfo {year} {2025})}\BibitemShut {NoStop}%
\bibitem [{\citenamefont {Morado}\ \emph {et~al.}(2025)\citenamefont {Morado}, \citenamefont {Zinovjev}, \citenamefont {Hedges}, \citenamefont {Cole},\ and\ \citenamefont {Michel}}]{morado2025enhancing}%
  \BibitemOpen
  \bibfield  {author} {\bibinfo {author} {\bibfnamefont {Jo{\~a}o}\ \bibnamefont {Morado}}, \bibinfo {author} {\bibfnamefont {Kirill}\ \bibnamefont {Zinovjev}}, \bibinfo {author} {\bibfnamefont {Lester~O}\ \bibnamefont {Hedges}}, \bibinfo {author} {\bibfnamefont {Daniel~J}\ \bibnamefont {Cole}}, \ and\ \bibinfo {author} {\bibfnamefont {Julien}\ \bibnamefont {Michel}},\ }\bibfield  {title} {\enquote {\bibinfo {title} {Enhancing electrostatic embedding for ml/mm free energy calculations},}\ }\href@noop {} {\bibfield  {journal} {\bibinfo  {journal} {Journal of Chemical Theory and Computation}\ } (\bibinfo {year} {2025})}\BibitemShut {NoStop}%
\bibitem [{\citenamefont {Marenich}\ \emph {et~al.}(2012)\citenamefont {Marenich}, \citenamefont {Jerome}, \citenamefont {Cramer},\ and\ \citenamefont {Truhlar}}]{marenich2012charge}%
  \BibitemOpen
  \bibfield  {author} {\bibinfo {author} {\bibfnamefont {Aleksandr~V}\ \bibnamefont {Marenich}}, \bibinfo {author} {\bibfnamefont {Steven~V}\ \bibnamefont {Jerome}}, \bibinfo {author} {\bibfnamefont {Christopher~J}\ \bibnamefont {Cramer}}, \ and\ \bibinfo {author} {\bibfnamefont {Donald~G}\ \bibnamefont {Truhlar}},\ }\bibfield  {title} {\enquote {\bibinfo {title} {Charge model 5: An extension of hirshfeld population analysis for the accurate description of molecular interactions in gaseous and condensed phases},}\ }\href@noop {} {\bibfield  {journal} {\bibinfo  {journal} {Journal of chemical theory and computation}\ }\textbf {\bibinfo {volume} {8}},\ \bibinfo {pages} {527--541} (\bibinfo {year} {2012})}\BibitemShut {NoStop}%
\bibitem [{\citenamefont {Wiberg}\ and\ \citenamefont {Rablen}(1993)}]{wiberg1993comparison}%
  \BibitemOpen
  \bibfield  {author} {\bibinfo {author} {\bibfnamefont {Kenneth~B}\ \bibnamefont {Wiberg}}\ and\ \bibinfo {author} {\bibfnamefont {Paul~R}\ \bibnamefont {Rablen}},\ }\bibfield  {title} {\enquote {\bibinfo {title} {Comparison of atomic charges derived via different procedures},}\ }\href@noop {} {\bibfield  {journal} {\bibinfo  {journal} {Journal of Computational Chemistry}\ }\textbf {\bibinfo {volume} {14}},\ \bibinfo {pages} {1504--1518} (\bibinfo {year} {1993})}\BibitemShut {NoStop}%
\bibitem [{\citenamefont {Mulliken}(1955)}]{mulliken1955electronic}%
  \BibitemOpen
  \bibfield  {author} {\bibinfo {author} {\bibfnamefont {Robert~S}\ \bibnamefont {Mulliken}},\ }\bibfield  {title} {\enquote {\bibinfo {title} {Electronic population analysis on lcao--mo molecular wave functions. i},}\ }\href@noop {} {\bibfield  {journal} {\bibinfo  {journal} {The Journal of chemical physics}\ }\textbf {\bibinfo {volume} {23}},\ \bibinfo {pages} {1833--1840} (\bibinfo {year} {1955})}\BibitemShut {NoStop}%
\bibitem [{\citenamefont {Hirshfeld}(1977)}]{hirshfeld1977bonded}%
  \BibitemOpen
  \bibfield  {author} {\bibinfo {author} {\bibfnamefont {Fred~L}\ \bibnamefont {Hirshfeld}},\ }\bibfield  {title} {\enquote {\bibinfo {title} {Bonded-atom fragments for describing molecular charge densities},}\ }\href@noop {} {\bibfield  {journal} {\bibinfo  {journal} {Theoretica chimica acta}\ }\textbf {\bibinfo {volume} {44}},\ \bibinfo {pages} {129--138} (\bibinfo {year} {1977})}\BibitemShut {NoStop}%
\bibitem [{\citenamefont {Verstraelen}\ \emph {et~al.}(2016)\citenamefont {Verstraelen}, \citenamefont {Vandenbrande}, \citenamefont {Heidar-Zadeh}, \citenamefont {Vanduyfhuys}, \citenamefont {Van~Speybroeck}, \citenamefont {Waroquier},\ and\ \citenamefont {Ayers}}]{verstraelen2016minimal}%
  \BibitemOpen
  \bibfield  {author} {\bibinfo {author} {\bibfnamefont {Toon}\ \bibnamefont {Verstraelen}}, \bibinfo {author} {\bibfnamefont {Steven}\ \bibnamefont {Vandenbrande}}, \bibinfo {author} {\bibfnamefont {Farnaz}\ \bibnamefont {Heidar-Zadeh}}, \bibinfo {author} {\bibfnamefont {Louis}\ \bibnamefont {Vanduyfhuys}}, \bibinfo {author} {\bibfnamefont {Veronique}\ \bibnamefont {Van~Speybroeck}}, \bibinfo {author} {\bibfnamefont {Michel}\ \bibnamefont {Waroquier}}, \ and\ \bibinfo {author} {\bibfnamefont {Paul~W}\ \bibnamefont {Ayers}},\ }\bibfield  {title} {\enquote {\bibinfo {title} {Minimal basis iterative stockholder: atoms in molecules for force-field development},}\ }\href@noop {} {\bibfield  {journal} {\bibinfo  {journal} {Journal of Chemical Theory and Computation}\ }\textbf {\bibinfo {volume} {12}},\ \bibinfo {pages} {3894--3912} (\bibinfo {year} {2016})}\BibitemShut {NoStop}%
\bibitem [{\citenamefont {Leontyev}\ \emph {et~al.}(2003)\citenamefont {Leontyev}, \citenamefont {Vener}, \citenamefont {Rostov}, \citenamefont {Basilevsky},\ and\ \citenamefont {Newton}}]{leontyev2003continuum}%
  \BibitemOpen
  \bibfield  {author} {\bibinfo {author} {\bibfnamefont {IV}~\bibnamefont {Leontyev}}, \bibinfo {author} {\bibfnamefont {MV}~\bibnamefont {Vener}}, \bibinfo {author} {\bibfnamefont {IV}~\bibnamefont {Rostov}}, \bibinfo {author} {\bibfnamefont {MV}~\bibnamefont {Basilevsky}}, \ and\ \bibinfo {author} {\bibfnamefont {Marshall~D}\ \bibnamefont {Newton}},\ }\bibfield  {title} {\enquote {\bibinfo {title} {Continuum level treatment of electronic polarization in the framework of molecular simulations of solvation effects},}\ }\href@noop {} {\bibfield  {journal} {\bibinfo  {journal} {The Journal of chemical physics}\ }\textbf {\bibinfo {volume} {119}},\ \bibinfo {pages} {8024--8037} (\bibinfo {year} {2003})}\BibitemShut {NoStop}%
\bibitem [{\citenamefont {Jorge}(2024)}]{jorge2024theoretically}%
  \BibitemOpen
  \bibfield  {author} {\bibinfo {author} {\bibfnamefont {Miguel}\ \bibnamefont {Jorge}},\ }\bibfield  {title} {\enquote {\bibinfo {title} {Theoretically grounded approaches to account for polarization effects in fixed-charge force fields},}\ }\href@noop {} {\bibfield  {journal} {\bibinfo  {journal} {The Journal of Chemical Physics}\ }\textbf {\bibinfo {volume} {161}} (\bibinfo {year} {2024})}\BibitemShut {NoStop}%
\bibitem [{\citenamefont {Drautz}(2019)}]{drautz2019atomic}%
  \BibitemOpen
  \bibfield  {author} {\bibinfo {author} {\bibfnamefont {Ralf}\ \bibnamefont {Drautz}},\ }\bibfield  {title} {\enquote {\bibinfo {title} {Atomic cluster expansion for accurate and transferable interatomic potentials},}\ }\href@noop {} {\bibfield  {journal} {\bibinfo  {journal} {Physical Review B}\ }\textbf {\bibinfo {volume} {99}},\ \bibinfo {pages} {014104} (\bibinfo {year} {2019})}\BibitemShut {NoStop}%
\bibitem [{\citenamefont {Kocer}\ \emph {et~al.}(2025)\citenamefont {Kocer}, \citenamefont {Singraber}, \citenamefont {Finkler}, \citenamefont {Misof}, \citenamefont {Ko}, \citenamefont {Dellago},\ and\ \citenamefont {Behler}}]{Kocer2025Iterative}%
  \BibitemOpen
  \bibfield  {author} {\bibinfo {author} {\bibfnamefont {Emir}\ \bibnamefont {Kocer}}, \bibinfo {author} {\bibfnamefont {Andreas}\ \bibnamefont {Singraber}}, \bibinfo {author} {\bibfnamefont {Jonas~A.}\ \bibnamefont {Finkler}}, \bibinfo {author} {\bibfnamefont {Philipp}\ \bibnamefont {Misof}}, \bibinfo {author} {\bibfnamefont {Tsz~Wai}\ \bibnamefont {Ko}}, \bibinfo {author} {\bibfnamefont {Christoph}\ \bibnamefont {Dellago}}, \ and\ \bibinfo {author} {\bibfnamefont {J{\"o}rg}\ \bibnamefont {Behler}},\ }\bibfield  {title} {\enquote {\bibinfo {title} {Iterative charge equilibration for fourth-generation high-dimensional neural network potentials},}\ }\href {\doibase 10.1063/5.0252566} {\bibfield  {journal} {\bibinfo  {journal} {The Journal of Chemical Physics}\ }\textbf {\bibinfo {volume} {162}},\ \bibinfo {pages} {124106} (\bibinfo {year} {2025})}\BibitemShut {NoStop}%
\bibitem [{\citenamefont {Gubler}\ \emph {et~al.}(2024)\citenamefont {Gubler}, \citenamefont {Finkler}, \citenamefont {Sch{\"a}fer}, \citenamefont {Behler},\ and\ \citenamefont {Goedecker}}]{Gubler2024Acceleratinga}%
  \BibitemOpen
  \bibfield  {author} {\bibinfo {author} {\bibfnamefont {Moritz}\ \bibnamefont {Gubler}}, \bibinfo {author} {\bibfnamefont {Jonas~A.}\ \bibnamefont {Finkler}}, \bibinfo {author} {\bibfnamefont {Moritz~R.}\ \bibnamefont {Sch{\"a}fer}}, \bibinfo {author} {\bibfnamefont {J{\"o}rg}\ \bibnamefont {Behler}}, \ and\ \bibinfo {author} {\bibfnamefont {Stefan}\ \bibnamefont {Goedecker}},\ }\bibfield  {title} {\enquote {\bibinfo {title} {Accelerating fourth-generation machine learning potentials using quasi-linear scaling particle mesh charge equilibration},}\ }\href {\doibase 10.1021/acs.jctc.4c00334} {\bibfield  {journal} {\bibinfo  {journal} {Journal of Chemical Theory and Computation}\ }\textbf {\bibinfo {volume} {20}},\ \bibinfo {pages} {7264--7271} (\bibinfo {year} {2024})}\BibitemShut {NoStop}%
\bibitem [{\citenamefont {Siepmann}\ and\ \citenamefont {Sprik}(1995)}]{siepmann1995influence}%
  \BibitemOpen
  \bibfield  {author} {\bibinfo {author} {\bibfnamefont {J~Ilja}\ \bibnamefont {Siepmann}}\ and\ \bibinfo {author} {\bibfnamefont {Michiel}\ \bibnamefont {Sprik}},\ }\bibfield  {title} {\enquote {\bibinfo {title} {Influence of surface topology and electrostatic potential on water/electrode systems},}\ }\href@noop {} {\bibfield  {journal} {\bibinfo  {journal} {The Journal of chemical physics}\ }\textbf {\bibinfo {volume} {102}},\ \bibinfo {pages} {511--524} (\bibinfo {year} {1995})}\BibitemShut {NoStop}%
\bibitem [{\citenamefont {Simeon}\ \emph {et~al.}(2025)\citenamefont {Simeon}, \citenamefont {Mirarchi}, \citenamefont {Pelaez}, \citenamefont {Galvelis},\ and\ \citenamefont {De~Fabritiis}}]{Simeon2025Broadening}%
  \BibitemOpen
  \bibfield  {author} {\bibinfo {author} {\bibfnamefont {Guillem}\ \bibnamefont {Simeon}}, \bibinfo {author} {\bibfnamefont {Antonio}\ \bibnamefont {Mirarchi}}, \bibinfo {author} {\bibfnamefont {Raul~P.}\ \bibnamefont {Pelaez}}, \bibinfo {author} {\bibfnamefont {Raimondas}\ \bibnamefont {Galvelis}}, \ and\ \bibinfo {author} {\bibfnamefont {Gianni}\ \bibnamefont {De~Fabritiis}},\ }\bibfield  {title} {\enquote {\bibinfo {title} {Broadening the scope of neural network potentials through direct inclusion of additional molecular attributes},}\ }\href {\doibase 10.1021/acs.jctc.4c01625} {\bibfield  {journal} {\bibinfo  {journal} {Journal of Chemical Theory and Computation}\ }\textbf {\bibinfo {volume} {21}},\ \bibinfo {pages} {1831--1837} (\bibinfo {year} {2025})}\BibitemShut {NoStop}%
\bibitem [{\citenamefont {Yuan}\ \emph {et~al.}(2025)\citenamefont {Yuan}, \citenamefont {Liu}, \citenamefont {Chen}, \citenamefont {Zhong}, \citenamefont {Raja}, \citenamefont {Kreiman}, \citenamefont {Vargas}, \citenamefont {Xu}, \citenamefont {{Head-Gordon}}, \citenamefont {Yang}, \citenamefont {Blau}, \citenamefont {Cheng}, \citenamefont {Krishnapriyan},\ and\ \citenamefont {{Head-Gordon}}}]{yuan2025foundation}%
  \BibitemOpen
  \bibfield  {author} {\bibinfo {author} {\bibfnamefont {Eric C.-Y.}\ \bibnamefont {Yuan}}, \bibinfo {author} {\bibfnamefont {Yunsheng}\ \bibnamefont {Liu}}, \bibinfo {author} {\bibfnamefont {Junmin}\ \bibnamefont {Chen}}, \bibinfo {author} {\bibfnamefont {Peichen}\ \bibnamefont {Zhong}}, \bibinfo {author} {\bibfnamefont {Sanjeev}\ \bibnamefont {Raja}}, \bibinfo {author} {\bibfnamefont {Tobias}\ \bibnamefont {Kreiman}}, \bibinfo {author} {\bibfnamefont {Santiago}\ \bibnamefont {Vargas}}, \bibinfo {author} {\bibfnamefont {Wenbin}\ \bibnamefont {Xu}}, \bibinfo {author} {\bibfnamefont {Martin}\ \bibnamefont {{Head-Gordon}}}, \bibinfo {author} {\bibfnamefont {Chao}\ \bibnamefont {Yang}}, \bibinfo {author} {\bibfnamefont {Samuel~M.}\ \bibnamefont {Blau}}, \bibinfo {author} {\bibfnamefont {Bingqing}\ \bibnamefont {Cheng}}, \bibinfo {author} {\bibfnamefont {Aditi}\ \bibnamefont {Krishnapriyan}}, \ and\ \bibinfo {author} {\bibfnamefont {Teresa}\ \bibnamefont {{Head-Gordon}}},\ }\bibfield  {title} {\enquote {\bibinfo
  {title} {Foundation models for atomistic simulation of chemistry and materials},}\ }\href@noop {} {\bibfield  {journal} {\bibinfo  {journal} {arXiv preprint arXiv:2503.10538}\ } (\bibinfo {year} {2025})}\BibitemShut {NoStop}%
\bibitem [{\citenamefont {Eastman}\ \emph {et~al.}(2023)\citenamefont {Eastman}, \citenamefont {Behara}, \citenamefont {Dotson}, \citenamefont {Galvelis}, \citenamefont {Herr}, \citenamefont {Horton}, \citenamefont {Mao}, \citenamefont {Chodera}, \citenamefont {Pritchard}, \citenamefont {Wang}, \citenamefont {De~Fabritiis},\ and\ \citenamefont {Markland}}]{eastman2023spice}%
  \BibitemOpen
  \bibfield  {author} {\bibinfo {author} {\bibfnamefont {Peter}\ \bibnamefont {Eastman}}, \bibinfo {author} {\bibfnamefont {Pavan~Kumar}\ \bibnamefont {Behara}}, \bibinfo {author} {\bibfnamefont {David~L.}\ \bibnamefont {Dotson}}, \bibinfo {author} {\bibfnamefont {Raimondas}\ \bibnamefont {Galvelis}}, \bibinfo {author} {\bibfnamefont {John~E.}\ \bibnamefont {Herr}}, \bibinfo {author} {\bibfnamefont {Josh~T.}\ \bibnamefont {Horton}}, \bibinfo {author} {\bibfnamefont {Yuezhi}\ \bibnamefont {Mao}}, \bibinfo {author} {\bibfnamefont {John~D.}\ \bibnamefont {Chodera}}, \bibinfo {author} {\bibfnamefont {Benjamin~P.}\ \bibnamefont {Pritchard}}, \bibinfo {author} {\bibfnamefont {Yuanqing}\ \bibnamefont {Wang}}, \bibinfo {author} {\bibfnamefont {Gianni}\ \bibnamefont {De~Fabritiis}}, \ and\ \bibinfo {author} {\bibfnamefont {Thomas~E}\ \bibnamefont {Markland}},\ }\bibfield  {title} {\enquote {\bibinfo {title} {Spice, a dataset of drug-like molecules and peptides for training machine learning potentials},}\ }\href@noop
  {} {\bibfield  {journal} {\bibinfo  {journal} {Scientific Data}\ }\textbf {\bibinfo {volume} {10}},\ \bibinfo {pages} {11} (\bibinfo {year} {2023})}\BibitemShut {NoStop}%
\bibitem [{\citenamefont {Kov{\'a}cs}\ \emph {et~al.}(2025)\citenamefont {Kov{\'a}cs}, \citenamefont {Moore}, \citenamefont {Browning}, \citenamefont {Batatia}, \citenamefont {Horton}, \citenamefont {Pu}, \citenamefont {Kapil}, \citenamefont {Witt}, \citenamefont {Magd{\u a}u}, \citenamefont {Cole},\ and\ \citenamefont {Cs{\'a}nyi}}]{kovacs2025MACEOFF}%
  \BibitemOpen
  \bibfield  {author} {\bibinfo {author} {\bibfnamefont {D{\'a}vid~P{\'e}ter}\ \bibnamefont {Kov{\'a}cs}}, \bibinfo {author} {\bibfnamefont {J.~Harry}\ \bibnamefont {Moore}}, \bibinfo {author} {\bibfnamefont {Nicholas~J.}\ \bibnamefont {Browning}}, \bibinfo {author} {\bibfnamefont {Ilyes}\ \bibnamefont {Batatia}}, \bibinfo {author} {\bibfnamefont {Joshua~T.}\ \bibnamefont {Horton}}, \bibinfo {author} {\bibfnamefont {Yixuan}\ \bibnamefont {Pu}}, \bibinfo {author} {\bibfnamefont {Venkat}\ \bibnamefont {Kapil}}, \bibinfo {author} {\bibfnamefont {William~C.}\ \bibnamefont {Witt}}, \bibinfo {author} {\bibfnamefont {Ioan-Bogdan}\ \bibnamefont {Magd{\u a}u}}, \bibinfo {author} {\bibfnamefont {Daniel~J.}\ \bibnamefont {Cole}}, \ and\ \bibinfo {author} {\bibfnamefont {G{\'a}bor}\ \bibnamefont {Cs{\'a}nyi}},\ }\bibfield  {title} {\enquote {\bibinfo {title} {Mace-off: Short-range transferable machine learning force fields for organic molecules},}\ }\href@noop {} {\bibfield  {journal} {\bibinfo  {journal} {Journal of the
  American Chemical Society}\ } (\bibinfo {year} {2025})}\BibitemShut {NoStop}%
\bibitem [{\citenamefont {Barroso-Luque}\ \emph {et~al.}(2024)\citenamefont {Barroso-Luque}, \citenamefont {Shuaibi}, \citenamefont {Fu}, \citenamefont {Wood}, \citenamefont {Dzamba}, \citenamefont {Gao}, \citenamefont {Rizvi}, \citenamefont {Zitnick},\ and\ \citenamefont {Ulissi}}]{barroso_omat24}%
  \BibitemOpen
  \bibfield  {author} {\bibinfo {author} {\bibfnamefont {Luis}\ \bibnamefont {Barroso-Luque}}, \bibinfo {author} {\bibfnamefont {Muhammed}\ \bibnamefont {Shuaibi}}, \bibinfo {author} {\bibfnamefont {Xiang}\ \bibnamefont {Fu}}, \bibinfo {author} {\bibfnamefont {Brandon~M.}\ \bibnamefont {Wood}}, \bibinfo {author} {\bibfnamefont {Misko}\ \bibnamefont {Dzamba}}, \bibinfo {author} {\bibfnamefont {Meng}\ \bibnamefont {Gao}}, \bibinfo {author} {\bibfnamefont {Ammar}\ \bibnamefont {Rizvi}}, \bibinfo {author} {\bibfnamefont {C.~Lawrence}\ \bibnamefont {Zitnick}}, \ and\ \bibinfo {author} {\bibfnamefont {Zachary~W.}\ \bibnamefont {Ulissi}},\ }\bibfield  {title} {\enquote {\bibinfo {title} {Open materials 2024 (omat24) inorganic materials dataset and models},}\ }\href@noop {} {\bibfield  {journal} {\bibinfo  {journal} {arXiv prerpint arXiv:2410.12771}\ } (\bibinfo {year} {2024})}\BibitemShut {NoStop}%
\bibitem [{\citenamefont {Levine}\ \emph {et~al.}(2025)\citenamefont {Levine}, \citenamefont {Shuaibi}, \citenamefont {Spotte-Smith}, \citenamefont {Taylor}, \citenamefont {Hasyim}, \citenamefont {Michel}, \citenamefont {Batatia}, \citenamefont {Csányi}, \citenamefont {Dzamba}, \citenamefont {Eastman}, \citenamefont {Frey}, \citenamefont {Fu}, \citenamefont {Gharakhanyan}, \citenamefont {Krishnapriyan}, \citenamefont {Rackers}, \citenamefont {Raja}, \citenamefont {Rizvi}, \citenamefont {Rosen}, \citenamefont {Ulissi}, \citenamefont {Vargas}, \citenamefont {Zitnick}, \citenamefont {Blau},\ and\ \citenamefont {Wood}}]{levine2025openmolecules2025omol25}%
  \BibitemOpen
  \bibfield  {author} {\bibinfo {author} {\bibfnamefont {Daniel~S.}\ \bibnamefont {Levine}}, \bibinfo {author} {\bibfnamefont {Muhammed}\ \bibnamefont {Shuaibi}}, \bibinfo {author} {\bibfnamefont {Evan Walter~Clark}\ \bibnamefont {Spotte-Smith}}, \bibinfo {author} {\bibfnamefont {Michael~G.}\ \bibnamefont {Taylor}}, \bibinfo {author} {\bibfnamefont {Muhammad~R.}\ \bibnamefont {Hasyim}}, \bibinfo {author} {\bibfnamefont {Kyle}\ \bibnamefont {Michel}}, \bibinfo {author} {\bibfnamefont {Ilyes}\ \bibnamefont {Batatia}}, \bibinfo {author} {\bibfnamefont {Gábor}\ \bibnamefont {Csányi}}, \bibinfo {author} {\bibfnamefont {Misko}\ \bibnamefont {Dzamba}}, \bibinfo {author} {\bibfnamefont {Peter}\ \bibnamefont {Eastman}}, \bibinfo {author} {\bibfnamefont {Nathan~C.}\ \bibnamefont {Frey}}, \bibinfo {author} {\bibfnamefont {Xiang}\ \bibnamefont {Fu}}, \bibinfo {author} {\bibfnamefont {Vahe}\ \bibnamefont {Gharakhanyan}}, \bibinfo {author} {\bibfnamefont {Aditi~S.}\ \bibnamefont {Krishnapriyan}}, \bibinfo {author}
  {\bibfnamefont {Joshua~A.}\ \bibnamefont {Rackers}}, \bibinfo {author} {\bibfnamefont {Sanjeev}\ \bibnamefont {Raja}}, \bibinfo {author} {\bibfnamefont {Ammar}\ \bibnamefont {Rizvi}}, \bibinfo {author} {\bibfnamefont {Andrew~S.}\ \bibnamefont {Rosen}}, \bibinfo {author} {\bibfnamefont {Zachary}\ \bibnamefont {Ulissi}}, \bibinfo {author} {\bibfnamefont {Santiago}\ \bibnamefont {Vargas}}, \bibinfo {author} {\bibfnamefont {C.~Lawrence}\ \bibnamefont {Zitnick}}, \bibinfo {author} {\bibfnamefont {Samuel~M.}\ \bibnamefont {Blau}}, \ and\ \bibinfo {author} {\bibfnamefont {Brandon~M.}\ \bibnamefont {Wood}},\ }\bibfield  {title} {\enquote {\bibinfo {title} {The open molecules 2025 (omol25) dataset, evaluations, and models},}\ }\href@noop {} {\bibfield  {journal} {\bibinfo  {journal} {arXiv preprint arXiv:2505.08762}\ } (\bibinfo {year} {2025})}\BibitemShut {NoStop}%
\bibitem [{\citenamefont {Cheng}\ \emph {et~al.}(2019)\citenamefont {Cheng}, \citenamefont {Engel}, \citenamefont {Behler}, \citenamefont {Dellago},\ and\ \citenamefont {Ceriotti}}]{cheng2019ab}%
  \BibitemOpen
  \bibfield  {author} {\bibinfo {author} {\bibfnamefont {Bingqing}\ \bibnamefont {Cheng}}, \bibinfo {author} {\bibfnamefont {Edgar~A}\ \bibnamefont {Engel}}, \bibinfo {author} {\bibfnamefont {J{\"o}rg}\ \bibnamefont {Behler}}, \bibinfo {author} {\bibfnamefont {Christoph}\ \bibnamefont {Dellago}}, \ and\ \bibinfo {author} {\bibfnamefont {Michele}\ \bibnamefont {Ceriotti}},\ }\bibfield  {title} {\enquote {\bibinfo {title} {Ab initio thermodynamics of liquid and solid water},}\ }\href@noop {} {\bibfield  {journal} {\bibinfo  {journal} {Proceedings of the National Academy of Sciences}\ }\textbf {\bibinfo {volume} {116}},\ \bibinfo {pages} {1110--1115} (\bibinfo {year} {2019})}\BibitemShut {NoStop}%
\bibitem [{\citenamefont {Cheng}\ \emph {et~al.}(2021)\citenamefont {Cheng}, \citenamefont {Bethkenhagen}, \citenamefont {Pickard},\ and\ \citenamefont {Hamel}}]{cheng2021phase}%
  \BibitemOpen
  \bibfield  {author} {\bibinfo {author} {\bibfnamefont {Bingqing}\ \bibnamefont {Cheng}}, \bibinfo {author} {\bibfnamefont {Mandy}\ \bibnamefont {Bethkenhagen}}, \bibinfo {author} {\bibfnamefont {Chris~J}\ \bibnamefont {Pickard}}, \ and\ \bibinfo {author} {\bibfnamefont {Sebastien}\ \bibnamefont {Hamel}},\ }\bibfield  {title} {\enquote {\bibinfo {title} {Phase behaviours of superionic water at planetary conditions},}\ }\href@noop {} {\bibfield  {journal} {\bibinfo  {journal} {Nature physics}\ }\textbf {\bibinfo {volume} {17}},\ \bibinfo {pages} {1228--1232} (\bibinfo {year} {2021})}\BibitemShut {NoStop}%
\bibitem [{\citenamefont {Cheng}\ \emph {et~al.}(2020)\citenamefont {Cheng}, \citenamefont {Mazzola}, \citenamefont {Pickard},\ and\ \citenamefont {Ceriotti}}]{cheng2020evidence}%
  \BibitemOpen
  \bibfield  {author} {\bibinfo {author} {\bibfnamefont {Bingqing}\ \bibnamefont {Cheng}}, \bibinfo {author} {\bibfnamefont {Guglielmo}\ \bibnamefont {Mazzola}}, \bibinfo {author} {\bibfnamefont {Chris~J}\ \bibnamefont {Pickard}}, \ and\ \bibinfo {author} {\bibfnamefont {Michele}\ \bibnamefont {Ceriotti}},\ }\bibfield  {title} {\enquote {\bibinfo {title} {Evidence for supercritical behaviour of high-pressure liquid hydrogen},}\ }\href@noop {} {\bibfield  {journal} {\bibinfo  {journal} {Nature}\ }\textbf {\bibinfo {volume} {585}},\ \bibinfo {pages} {217--220} (\bibinfo {year} {2020})}\BibitemShut {NoStop}%
\bibitem [{\citenamefont {Yue}\ \emph {et~al.}(2021)\citenamefont {Yue}, \citenamefont {Muniz}, \citenamefont {Calegari~Andrade}, \citenamefont {Zhang}, \citenamefont {Car},\ and\ \citenamefont {Panagiotopoulos}}]{yue2021short}%
  \BibitemOpen
  \bibfield  {author} {\bibinfo {author} {\bibfnamefont {Shuwen}\ \bibnamefont {Yue}}, \bibinfo {author} {\bibfnamefont {Maria~Carolina}\ \bibnamefont {Muniz}}, \bibinfo {author} {\bibfnamefont {Marcos~F}\ \bibnamefont {Calegari~Andrade}}, \bibinfo {author} {\bibfnamefont {Linfeng}\ \bibnamefont {Zhang}}, \bibinfo {author} {\bibfnamefont {Roberto}\ \bibnamefont {Car}}, \ and\ \bibinfo {author} {\bibfnamefont {Athanassios~Z}\ \bibnamefont {Panagiotopoulos}},\ }\bibfield  {title} {\enquote {\bibinfo {title} {When do short-range atomistic machine-learning models fall short?}}\ }\href {\doibase 10.1063/5.0031215} {\bibfield  {journal} {\bibinfo  {journal} {J. Chem. Phys.}\ }\textbf {\bibinfo {volume} {154}},\ \bibinfo {pages} {034111} (\bibinfo {year} {2021})}\BibitemShut {NoStop}%
\bibitem [{\citenamefont {Batatia}\ \emph {et~al.}(2025)\citenamefont {Batatia}, \citenamefont {Batzner}, \citenamefont {Kov{\'a}cs}, \citenamefont {Musaelian}, \citenamefont {Simm}, \citenamefont {Drautz}, \citenamefont {Ortner}, \citenamefont {Kozinsky},\ and\ \citenamefont {Cs{\'a}nyi}}]{Batatia2025designa}%
  \BibitemOpen
  \bibfield  {author} {\bibinfo {author} {\bibfnamefont {Ilyes}\ \bibnamefont {Batatia}}, \bibinfo {author} {\bibfnamefont {Simon}\ \bibnamefont {Batzner}}, \bibinfo {author} {\bibfnamefont {D{\'a}vid~P{\'e}ter}\ \bibnamefont {Kov{\'a}cs}}, \bibinfo {author} {\bibfnamefont {Albert}\ \bibnamefont {Musaelian}}, \bibinfo {author} {\bibfnamefont {Gregor N.~C.}\ \bibnamefont {Simm}}, \bibinfo {author} {\bibfnamefont {Ralf}\ \bibnamefont {Drautz}}, \bibinfo {author} {\bibfnamefont {Christoph}\ \bibnamefont {Ortner}}, \bibinfo {author} {\bibfnamefont {Boris}\ \bibnamefont {Kozinsky}}, \ and\ \bibinfo {author} {\bibfnamefont {G{\'a}bor}\ \bibnamefont {Cs{\'a}nyi}},\ }\bibfield  {title} {\enquote {\bibinfo {title} {The design space of e(3)-equivariant atom-centred interatomic potentials},}\ }\href {\doibase 10.1038/s42256-024-00956-x} {\bibfield  {journal} {\bibinfo  {journal} {Nature Machine Intelligence}\ }\textbf {\bibinfo {volume} {7}},\ \bibinfo {pages} {56--67} (\bibinfo {year} {2025})}\BibitemShut {NoStop}%
\end{thebibliography}
%

\end{document}